\documentclass[floatfix,reprint,pra,aps,twocolumn,groupedaddress,amsmath,amssymb,footinbib]{revtex4-2}
\usepackage{xcolor}
\usepackage{graphicx}  
\usepackage{dcolumn}   
\usepackage{bm}        
\usepackage{verbatim}   
\usepackage[mathscr]{euscript}
\usepackage{epstopdf}
\usepackage{float}
\usepackage{ragged2e}
\usepackage{soul}
\usepackage{hyperref}
\hypersetup{
    colorlinks = true,
    linkcolor=blue
}
\colorlet{Mycolor1}{green!10!orange!90!}
\definecolor{mint}{rgb}{0.85,1,0.85}
\definecolor{dgreen}{rgb}{0,0.2,0}
\definecolor{RED}{rgb}{1,0,0}

\usepackage[normalem]{ulem}

\newcommand{\oldsection}[2]{\section*{{\color{RED}\rule[2.5pt]{#1mm}{1pt}}\kern-#1mm #2}}
\newcommand{\oldsubsection}[2]{\subsection*{{\color{RED}\rule[2.5pt]{#1mm}{1pt}}\kern-#1mm #2}}

\newcommand{\bea}{\begin{eqnarray}}
\newcommand{\eea}{\end{eqnarray}}

\newcommand{\ket}[1]{|#1\rangle}
\newcommand{\bra}[1]{\langle#1|}
\newcommand{\dn}{{\downarrow}}
\newcommand{\up}{{\uparrow}}
\newcommand{\updn}{{\updownarrow}}
\newcommand{\ip}[1]{{\langle #1 \rangle}}
\newcommand{\eq}[1]{Eq.~(\ref{#1})}
\newcommand{\eqs}[1]{Eqs.~(\ref{#1})}
\newcommand{\eqr}[1]{(\ref{#1})}

\newcommand\rout{\bgroup\markoverwith{\textcolor{red}{\rule[0.5ex]{2pt}{0.8pt}}}\ULon}

\newcommand\routh{\bgroup\markoverwith{\textcolor{magenta}{\rule[0.5ex]{2pt}{0.8pt}}}\ULon}

\newcommand{\R}{{(R)}}
\newcommand{\M}{{(M)}}

\newcommand{\T}{{(T)}}
\newcommand{\X}{{(X)}}
\newcommand{\inl}{{\rm i}}
\newcommand{\fnl}{{\rm f}}
\newcommand{\x}{{(x)}}
\newcommand{\n}{{\overline{n}}}
\newcommand{\N}{{\overline{N}}}
\renewcommand{\L}{\mathcal{L}_{\rm s}}
\newcommand{\Q}{\mathcal{Q}_{\rm s}}
\newcommand{\DJz}{\Delta J_{z}}
\renewcommand{\P}{\mathcal{P}}
\newcommand{\sfc}[2]{\mbox{$\frac{#1}{#2}$}}
\newcommand{\ary}[2]{\begin{array}{#1} #2 \end{array}}
\renewcommand{\prd}{\operatorname*{\mbox{$\prod$}}}
\newcommand{\Var}{\text{Var}}

\newcounter{xx}

\pdfoutput=1

\begin{document}
\sethlcolor{mint}
\title{Thermodynamics of memory erasure via a spin reservoir}

\author{{T.} Croucher}
\affiliation{
   Centre for Quantum Dynamics,\\
   Griffith University,\\
   Brisbane, QLD 4111 Australia
   }
   \author{J.A. Vaccaro}
\affiliation{
   Centre for Quantum Dynamics,\\
   Griffith University,\\
   Brisbane, QLD 4111 Australia
   }
\date{\today}

\begin{abstract}
Thermodynamics with multiple-conserved quantities offers a promising direction for designing novel devices. For example, Vaccaro and Barnett's  [J. A. Vaccaro and S. M. Barnett, Proc. R. Soc. A 467, 1770 (2011); S. M. Barnett and J. A. Vaccaro, Entropy 15, 4956 (2013)] proposed information erasure scheme, where the cost of erasure is solely in terms of a conserved quantity other than energy, allows for new kinds of heat engines. In recent work, we studied the discrete fluctuations and average bounds of the erasure cost in spin angular momentum. Here we clarify the costs in terms of the spin equivalent of work, called spinlabor, and the spin equivalent of heat, called spintherm. We show that the previously-found bound on the erasure cost of $\gamma^{-1}\ln{2}$ can be violated by the spinlabor cost, and only applies to the spintherm cost. We obtain three bounds for spinlabor for different erasure protocols and determine the one that provides the tightest bound. For completeness, we derive a generalized Jarzynski equality and probability of violation which shows that for particular protocols the probability of violation can be surprisingly large. We also derive an integral fluctuation theorem and use it to analyze the cost of information erasure using a spin reservoir. 

\end{abstract}

\maketitle
\section{Introduction}
Landauer's erasure principle is essential to thermodynamics and information theory \cite{Landauer1961}. The principle sets a lower bound on the amount of work $W$ required to erase one bit of information as  $W\geq\beta^{-1}\ln2$, where $\beta$ is inverse  temperature of the surrounding environment \cite{Bennett1987}. Sagawa and Ueda \cite{Sagawa2009} showed that the average cost of erasing one bit of information can be less than allowed by Landauer's principle if the phase space volumes for each of the memory states are different. Nevertheless when erasure and measurement costs are combined, the overall cost satisfies Landauer's  bound. Gavrilov and Bechhoefer \cite{Bechhoefer2016} reconfirmed that violations of Landauer's principle for a memory consisting of an asymmetric double well potential are possible. They concluded that whether there is or is not a violation is a matter of semantics due to the non-equilibrium starting conditions of the system.

For the study of nanoscale systems \cite{Alhambra2016,Sagawa2012} where thermal fluctuations are important, violations of Landauer's principle are not a matter of semantics. In these particular systems, thermal fluctuations can reduce the erasure cost to below the bound given by Landauer's principle for a single shot. The cost averaged over all shots is, however, consistent with Landauer's principle. Dillenschneider and Lutz \cite{Dillenschneider2009} analyzed these fluctuations and obtained a bound for the probability of violation as
\bea
P(W \leq \beta ^{-1}\ln2-\epsilon) \leq e^{-\beta \epsilon},
\eea
where $P(W \leq \beta ^{-1}\ln2-\epsilon) $ is  the probability that the work $W$ required to erase 1 bit of entropy will be less than Landauer's bound of $\beta^{-1}\ln2$ an amount $\epsilon$.

Vaccaro and Barnett \cite{Vaccaro2011,Barnett2013}, were able to go beyond Landauer's principle to argue, using Jaynes maximum entropy principle \cite{Jaynes1957a,Jaynes1957b}, that information can be erased using arbitrary conserved quantities and that erasure need not incur an energy cost. They gave an explicit example showing that the erasure cost can be solely achieved in terms of spin-angular momentum when the erasure process makes use of an energy degenerate spin reservoir. In this case the erasure cost is given by
\bea
    \Delta J'_{z} \geq \gamma^{-1}\ln{2} \label{eqn:VB}
\eea
in terms of a change in spin angular momentum $J'_{z}$ where $\gamma$ is a Lagrange multiplier 
\bea
      \gamma=\frac{1}{\hbar}\ln\left[\frac{N\hbar-2\langle \hat{J}^{(R)}_{z}\rangle}{N\hbar+2\langle \hat{J}^{(R)}_{z}\rangle}\right]=\frac{1}{\hbar}\ln\left[\frac{1-\alpha}{\alpha}\right],
      \label{eqn:gamma}
\eea
the superscript ($R$) indicates the reservoir, $\langle\hat{J}^{(R)}_z\rangle=\left(\alpha-\frac{1}{2}\right)N\hbar$ is the $z$ component of the reservoir spin angular momentum, $N$ is the number of spins in the reservoir and $\alpha$ represents the spin polarisation parameter bounded such that $0\le \alpha \le 1$. Here we further restrict $\alpha$ to $0\le \alpha \le 0.5$ as this provides us with positive values of $\gamma$ which we refer to as inverse ``spin temperature''.

The novelty of Vaccaro and Barnett's discovery allows for new kinds of heat engines and batteries that use multiple conserved quantities. Work in this field has developed methods on how multiple conserved quantities can be extracted and stored into batteries with a trade-off between the conserved quantities in affect \cite{Guryanova2016}. Hybrid thermal machines, machines that can cool, heat and/or produce work simultaneously have been also extended into this new regime \cite{Manzano2020}. Other research has looked into generalised heat engines and batteries using a finite-size baths of multiple conserved quantities \cite{Ito2016}. Furthermore a quantum heat engine using a thermal and spin reservoir was proposed that produces no waste heat \cite{Wright2018, Croucher2018}.

In our recent Letter \cite{Croucher2017}, we stated an analogous first law of thermodynamics in terms of the conserved spin angular momentum,
\begin{align}    \label{eqn:first_law}
    \Delta J_{z}= \mathcal{L}_{\rm s}+\mathcal{Q}_{\rm s}
\end{align}
where
\begin{align}    \label{eqn:L_s}
     \mathcal{L}_{\rm s}=\sum_{j,m_j}\hbar p(j,m_j)\Delta g(m_j)
\end{align}
is the spinlabor (i.e. the spin equivalent of work) and
\begin{align}    \label{eqn:Q_s}
     \mathcal{Q}_{\rm s}=\sum_{j,m_j}\hbar g(m_j)\Delta p(j,m_j)
\end{align}
is the spintherm (i.e. the spin equivalent of heat), $p(j,m_j)$ is the probability associated with the occupation of the spin state $(j,m_{j})$, $g(m_j)=m_j$, and $j$ and $m_j$ are the usual angular momentum quantum numbers \cite{Croucher2017}. The authors of \cite{Wright2018, Croucher2018} have used spintherm and spinlabor in conjunction with the conventional heat and work resources in the design a \emph{spin heat engine} (SHE) that operates between a thermal and a spin reservoir. It's principle operation is to extract heat from the thermal reservoir and convert it into work as the output through dissipating spinlabor as spintherm in the spin reservoir. This necessity of spintherm production within the model represents an alternate resolution of the Maxwell-demon paradox \cite{Landauer1961, Bennett1987}, and so \eqr{eqn:VB} is equivalent to a statement of the second law for conservation of spin.

We also analyzed the fluctuations for the Vaccaro and Barnett (VB) erasure protocol and obtained the probability of violating the bound in \eq{eqn:VB}
\bea
Pr(\mathcal{L}_s \leq \gamma^{-1} \ln 2-\epsilon) &\leq& A e^{-\gamma \epsilon }
\eea
where $A \equiv \left( 1+e^{-\gamma\hbar}\right) \left( 1+e^{-2\gamma\hbar}\right)^{-1}$. We found a tighter, semi-analytical bound on the probability of violation given by
\bea
Pr(\mathcal{L}_s \leq \gamma^{-1} \ln 2-\epsilon) &\leq& A e^{-\sqrt{\frac{\gamma}{\hbar}} \epsilon },
\eea
in the limit as $\gamma$ approaches $0$.

In this work, we review the VB erasure protocol and then we generalize it to include variations \S\ref{sec:review}. In \S\ref{sec:analysis} we derive the spinlabor statistics associated with the protocol variations. We also derive the associated Jarzynski equality and find its corresponding probability of violation in \S\ref{sec:Jarzynski-like equality}. We include an analysis of the situation when the information stored in the memory is not maximal. In \S\ref{sec:integral fluctuation theorem} we derive an integral fluctuation theorem associated with spin reservoirs. We compare in \S\ref{sec:bounds} different bounds on the spinlabor and spintherm costs and determine the optimum. In \S\ref{sec:conclusion} we conclude by summarizing major results within the paper.

\section{Details of the erasure protocol}\label{sec:review}

\subsection{Review of the standard erasure protocol}
This section reviews the standard protocol analyzed in Ref \cite{Vaccaro2011,Barnett2013,Croucher2017}. The memory is a two-state system which is in contact with an energy-degenerate spin reservoir. The logic states of the memory are associated with the eigenstates of the $z$ component of spin polarization. These states are assumed to be energy degenerate to ensure that the erasure process incurs no energy cost. We also assume any spatial degrees of freedom do not play an active role in the erasure process and are traced over allowing us to focus exclusively on the spin degree of freedom.

The reservoir contains a very large number, $N$, of spin-$\frac{1}{2}$ particles in equilibrium at inverse spin temperature $\gamma$.
The memory spin is initially in the spin-down state (logical 0) with probability $p_{\dn}$ and spin-up (logical 1) with probability $p_\up=1-p_{\dn}$.
The reservoir has a probability distribution given by
\bea
P_{\up}(n)=\sum_{\nu=1}^{^{N}C_{n}}P_{\up}(n,\nu)=\sum_{\nu=1}^{^{N}C_{n}}\frac{e^{-\gamma n\hbar}}{Z_R\; ^{N}C_{n}}=\frac{e^{-\gamma n\hbar}}{Z_R}, \label{eqn:probr}
\eea
where $n$ is the number of spins in the spin-up state $\ket{\up}\bra{\up}$,  $\nu=1, 2,\ldots {}^{N}C_{n}$ indexes different states with the same value of $n$ and $Z_{R}$ is the associated partition function.
`The reservoir is used during the erasure process to absorb the unwanted entropy in the memory aided by ancillary spins that acts as a catalyst. The spin exchange between the memory, ancillary spins and the reservoir is assumed to conserve total spin, i.e. $\langle \Delta J_{z} \rangle=0$,  and will be the forum in which erasure occurs. The large number of spins in the reservoir compared to the single spin in the memory implies that the spin temperature of the reservoir remains approximately constant during the spin exchanges. At the conclusion of the erasure process, the ancillary spins are left in their initial state.

The process of erasure requires an energy degenerate ancillary spin-$\frac{1}{2}$ particle to be added to the memory.
This ancilla is initially in a state $|{\downarrow}\rangle\langle{\downarrow}|$ corresponding to the logical 0 state.
A controlled-not (CNOT) operation is applied to the memory-ancilla system with the memory spin acting as the control and the ancilla the target.
The applied CNOT operation leaves both memory and ancilla spins in the state $\ket{\up}\bra{\up}$  with probability $p_{\uparrow}$ and the state $\ket{\dn}\bra{\dn}$ with probability $1-p_{\uparrow}$.
Following the application of the CNOT operation, the memory-ancilla system is allowed to reach spin equilibrium with the reservoir through the exchange of angular momentum in multiples of $2\hbar$ between the memory-ancilla system and random pairs of spins in the reservoir.
This equilibration step conserves spin angular momentum and is where entropy is removed from the memory spin; it treats the memory-ancilla system as effectively being a 2 state system where all memory-ancilla spins are correlated and in the same spin state (i.e.
the only possibilities are that all spins are spin-up or all are spin-down). An erasure cycle of adding an ancilla to the memory-ancilla system, applying a CNOT operation, and spin equilibration through the exchange of fixed multiples of $\hbar$ with the spin reservoir is repeated indefinitely, in principle.

For later reference, the combined process of adding an ancilla and performing the CNOT operation on the memory-ancilla system will be called simply a \textit{CNOT step} and, separately, the equilibration between the memory-ancilla system with the spin reservoir will be called the \textit{equilibration step}, for convenience.

\subsection{Variations}\label{sec:3 protocols}

The protocol just described, comprising of an alternating sequence of CNOT and equilibration steps beginning with a CNOT step, is the \emph{standard} one that was introduced by Vaccaro and Barnett \cite{Vaccaro2011} and has been used elsewhere \cite{Barnett2013,Croucher2017}.
Variations arise when the sequence of steps is permuted.
For example, instead of the erasure process beginning with a CNOT step, it could begin with an equilibration step and continue with the regular CNOT-equilibration cycles.
Alternatively, a number of CNOT steps could be applied before the first equilibration step, and so on.
When considering various orderings two points immediately come to mind.
The first is that a sequence of equilibration steps is equivalent, in resource terms, to a single equilibration step as the memory, ancilla and reservoir is not changed statistically after the first one, and so we needn't consider them further.
In contrast, a sequence of CNOT steps is markedly different from a single CNOT step if the memory-ancilla system is in the $\ket{\up}\bra{\up}$, as each one incurs a spinlabor cost of $1\hbar$.
The second point is that beginning the erasure process with an equilibration step will remove all evidence of the initial state of the memory and replace its initial probabilities $p_\up$ and $p_\dn=1-p_\up$ of being in the states $\ket{\up}\bra{\up}$ and $\ket{\dn}\bra{\dn}$, respectively, with corresponding probabilities associated with the spin reservoir, and so the subsequent spinlabor cost of the erasure will, therefore, be independent of the initial contents of the memory.

\begin{figure*}
	\centering
	\includegraphics[width=\textwidth]{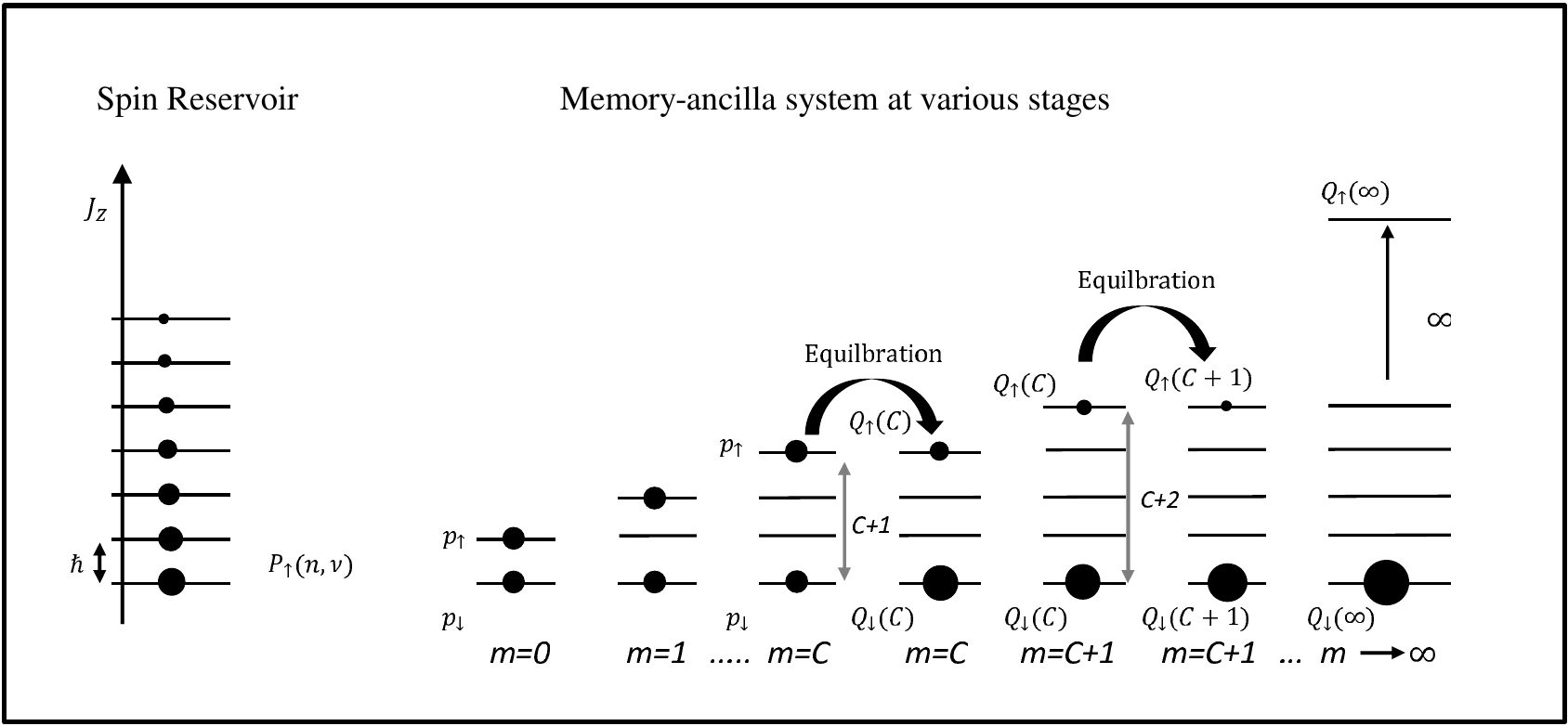}
	\vspace{-2mm}
	\caption{An illustration of the erasure process for an arbitrary protocol. The upwards vertical direction represents increasing values of the $z$ component of angular momentum. The state of the spin reservoir is represented on the far left by a spin level diagram. The remaining spin level diagrams to its right represent the spin state of the memory-ancilla system at various stages of the erasure process. The value of $m$ is the number of CNOT steps that have taken place. The illustration shows the specific case of $C=2$, where there are two probabilities at $m=C$, one before equilibration and one after equilibration. Other values are $p_{\uparrow}=0.5$, $p_{\downarrow}=1-p_{\uparrow}$, $Q_\uparrow(m) =\frac{e^{-(m+1)\gamma\hbar}}{1+e^{-(m+1)\gamma\hbar}}$ for $m\geq 0$ and $Q_\downarrow(m)=1-Q_\uparrow(m)$ for $m\geq 0$.}
	\label{fig:Erasure_process}
	\vspace{-2mm}
\end{figure*}

We wish to investigate the consequences of variations at the start of the erasure process.  Accordingly, we define the variable $C$ to be the number of CNOT steps that are applied before the first equilibration step, after which the regular cycles comprising of a CNOT step followed by an equilibration step are applied, as in the standard protocol.
This means that the value of $C$ indicates the nature of the variation in the erasure protocol, with $C=1$ corresponding to the standard protocol.  Also, to keep track of the position in the sequence of steps, we define the variable $m$ to be the number of CNOT steps that have been performed.  Every variant of the erasure protocol begins with $m=0$ corresponding to the initial state of the memory.  Figure \ref{fig:Erasure_process} illustrates the values of $C$ and $m$ for an arbitrary protocol with $C>0$.

\section{Statistics of the erasure costs \label{sec:analysis}}

In this section, we analyse the spinlabor and spintherm costs for a generic protocol. Unless it is clear from the context, we will differentiate the cost that accumulates over multiple steps from that of a single step by qualifying the former as the \emph{accumulated} cost, as in the accumulated spinlabor cost and the accumulated spintherm cost.

\subsection{Spinlabor statistics \label{sec:spinlabor}}

The CNOT operation incurs a spinlabor cost of $\hbar$ when the memory is in the $\ket{\up}\bra{\up}$ state.  Initially, the average cost of the operation is $p_{\uparrow}\hbar$ where $p_{\uparrow}$ is the initial probability that the memory is in this state.  If $C$ CNOT operations are performed before the first equilibration step, then the average of the accumulated spinlabor cost incurred is $ C p_{\uparrow}\hbar$.

Each time an equilibration step is performed, it leaves the memory-ancilla system in a statistical state that is uncorrelated to what it was prior to the step.  Let $Q_{\up}(m)$ be the probability that the memory-ancilla spins are all in the $\ket{\up}\bra{\up}$ state just after an equilibration step for the general case where $m$ prior CNOT operations have been performed. The equilibration process randomly exchanges spin-angular momentum between the reservoir and the memory-ancilla system in multiples of $(m+1)\hbar$, and so $Q_{\up}(m)$ becomes equal to the corresponding relative probability for the reservoir, and so \cite{Vaccaro2011,Barnett2013}
\bea
      Q_\up(m) =\frac{P_{\up}(m+1)}{P_{\up}(0)+P_{\up}(m+1)}=\frac{e^{-(m+1)\gamma\hbar}}{1+e^{-(m+1)\gamma\hbar}}  \label{eqn:Q}
\eea
and $Q_\downarrow(m)=1-Q_\uparrow(m)$, where $P_{\up}(m)$ is given by \eq{eqn:probr}. In the case of the first equilibration step, $m=C$. The memory is partially erased if the probability of the memory being in the spin up state is reduced during an equilibration step.

The average spinlabor cost of a subsequent CNOT step is $\hbar Q_\uparrow(C)$. Thus performing further cycles comprising of an equilibration step followed by an ancilla addition-CNOT operation gives additional average costs of $\hbar Q_{\up}(C+1)$,  $\hbar Q_{\up}(C+2)$ and so on.

Combining the costs before, $\hbar C  p_{\uparrow}$, and after, $\sum_{m=C}^{\infty} \hbar Q_\uparrow(m)$, the first equilibration step gives the average accumulated spinlabor cost as
\bea
    \langle\mathcal{L}_{\rm s}\rangle_C=\hbar C p_{\uparrow}+\sum_{m=C}^{\infty} \hbar Q_\uparrow(m). \label{eqn:avg}
\eea
The subscript on the left side indicates the dependence of the expectation value $\ip{\cdot}_C$ on the protocol variation parameter $C$.

We now examine the \emph{fluctuations} in the accumulated spinlabor cost for an erasure protocol for an arbitrary value of $C$.
We need to keep track of the number of CNOT steps as the spinlabor cost accumulates, and so we introduce a more concise notation. Let $\P_m(n)$ be the probability that the accumulative spinlabor cost is $\L=n \hbar$ after  $m$ CNOT operations have been performed.
Clearly $n$ cannot exceed the number of CNOT operations nor can it be negative, and so $\P_m(n)=0$ unless $0 \leq n \leq m$.
The end of the erasure process corresponds to the limit $m\to\infty$ and so the probability that an erasure protocol will incur a spinlabor cost of $\L$ is given by
\begin{align}    \label{eqn:Pr(L)=P_infty(n)}
     Pr(\L)=\P_\infty(n) \text{ for } \L=n \hbar.
\end{align}
The initial values of $\P_m(n)$ before anything is done (i.e. for $m=0$) are simply
\begin{align}
\P_{0}(n)&=\left\{\ary{l}{1, \text{ for }n=0\\
                          0, \text{ otherwise},
                          }\right. \label{eqn:secondeq m=0}
\end{align}
that is, initially the accumulated spinlabor cost is zero.
Each CNOT operation contributes a cost of $\hbar$ with the probability of either $p_{\up}$ before the first equilibration step, or $Q_{\up}(m)$ given in Eq. \eqref{eqn:Q} after it.

Before the first equilibration step, the spinlabor cost after $m$ CNOT operations is $m\hbar$ with probability $p_\up$ and $0$ with probability $p_\dn=1-p_\up$. The probability $\P_m(n)$ is therefore given by
\bea
\P_{m}(0)&=& 1-p_{\uparrow} \nonumber \\
\P_{m}(m)&=& p_{\uparrow} \label{eqn:secondeq}
\eea
and $\P_m(n)=0$ for $n=1,2,\ldots{}, m - 1$ and $0<m \le C$.

We calculate the probability $\P_{m}(n)$ for $m>C$, i.e. for CNOT steps after the first equilibration step has occurred, by considering the possibilities for the cost previously being $n\hbar$ and not increasing, and previously being $(n-1)\hbar$ and increasing by $1\hbar$, i.e. $\P_{m}(n)$ is given by
\bea
 &  Pr\left(\kern-1mm\begin{array}{c}
    \text{previous}\\ \text{cost is } n\hbar
   \end{array}\kern-1mm\right)
     \times
    Pr\left(\kern-1mm\begin{array}{c}
     \text{memory is}\\ \text{spin-down}
  \end{array}\kern-1mm\right)\notag\\
&\qquad +
Pr\left(\kern-1mm\begin{array}{c}
    \text{previous}\\ \text{cost is } (n-1)\hbar
   \end{array}\kern-1mm\right)
    \times
     Pr\left(\kern-1mm\begin{array}{c}
    \text{memory is}\\ \text{spin-up}
  \end{array}\kern-1mm\right),\notag
\eea
where $Pr(X)$ represents the probability of $X$.
Recalling Eq. \eqref{eqn:Q}, this yields the recurrence relation
\bea
    \P_{m}(n)&=&\P_{m-1}(n)Q_{\dn} (m-1)\nonumber\\&&\quad +\P_{m-1}(n-1)Q_{\up} (m-1), \label{eqn:recrel}
\eea
for $m>C$, where we set $\P_{m}(n)=0$ for $n<0$ for convenience.
The statistics of a complete erasure process are obtained in the $m\to\infty$ limit.
We derive analytic solutions of this recurrence relation in \hyperref[sec:ap anal prob soln]{Appendix \ref{sec:ap anal prob soln}}.
Keeping in mind the change of notation in \eq{eqn:Pr(L)=P_infty(n)}, the probability that the spinlabor cost is $\L=n\hbar$ for the case $C=0$, where an equilibration step occurs before the first CNOT step, is shown by \eq{eqn:ap P_infty for C=0} to be
\begin{align}    \label{eqn:P_infty for C=0}
        Pr(\L)=\frac{ e^{-\frac{1}{2}n(n+1)\gamma\hbar}}{(e^{-\gamma\hbar};e^{-\gamma\hbar})_n(-e^{-\gamma \hbar};e^{-\gamma \hbar})_\infty},
\end{align}
and for the case  $C>0$, where $C$ CNOT steps occur before the first equilibration step, is shown by \eq{eqn:ap P_infty for C>0} to be
\begin{align}   \label{eqn:P_infty for C>0, n<C}
    Pr(\L)&=
        p_\dn\frac{e^{-n(C+\frac{n+1}{2})\gamma\hbar}}{(e^{-\gamma\hbar};e^{-\gamma\hbar})_n(-e^{-\gamma \hbar};e^{-\gamma \hbar})_\infty}
\end{align}
for $n<C $ and
\begin{align}   \label{eqn:P_infty for C>0, n>C}
    Pr(\L)&=
        p_\dn\frac{e^{-n(C+\frac{n+1}{2})\gamma\hbar}}{(e^{-\gamma\hbar};e^{-\gamma\hbar})_n(-e^{-\gamma \hbar};e^{-\gamma \hbar})_\infty}
           \notag\\
        &\qquad + p_\up \frac{ e^{-(n-C)(C+\frac{n-C+1}{2})\gamma\hbar}}{(e^{-\gamma\hbar};e^{-\gamma\hbar})_{n-C}(-e^{-\gamma \hbar};e^{-\gamma \hbar})_\infty}
\end{align}
for $n\ge C$, where $(a;q)_n\equiv\prod_{k=0}^{n-1}(1-a q^k)$ is the $q$-Pochhammer symbol.
Substituting $C=0$ into \eq{eqn:P_infty for C>0, n>C} and using $p_\up+p_\dn=1$ gives the same result as \eq{eqn:P_infty for C=0} and confirms our expectation that the $C=0$ protocol is independent of the initial contents of the memory.

\begin{figure}[H]
	\centering
	\includegraphics[width=0.48\textwidth]{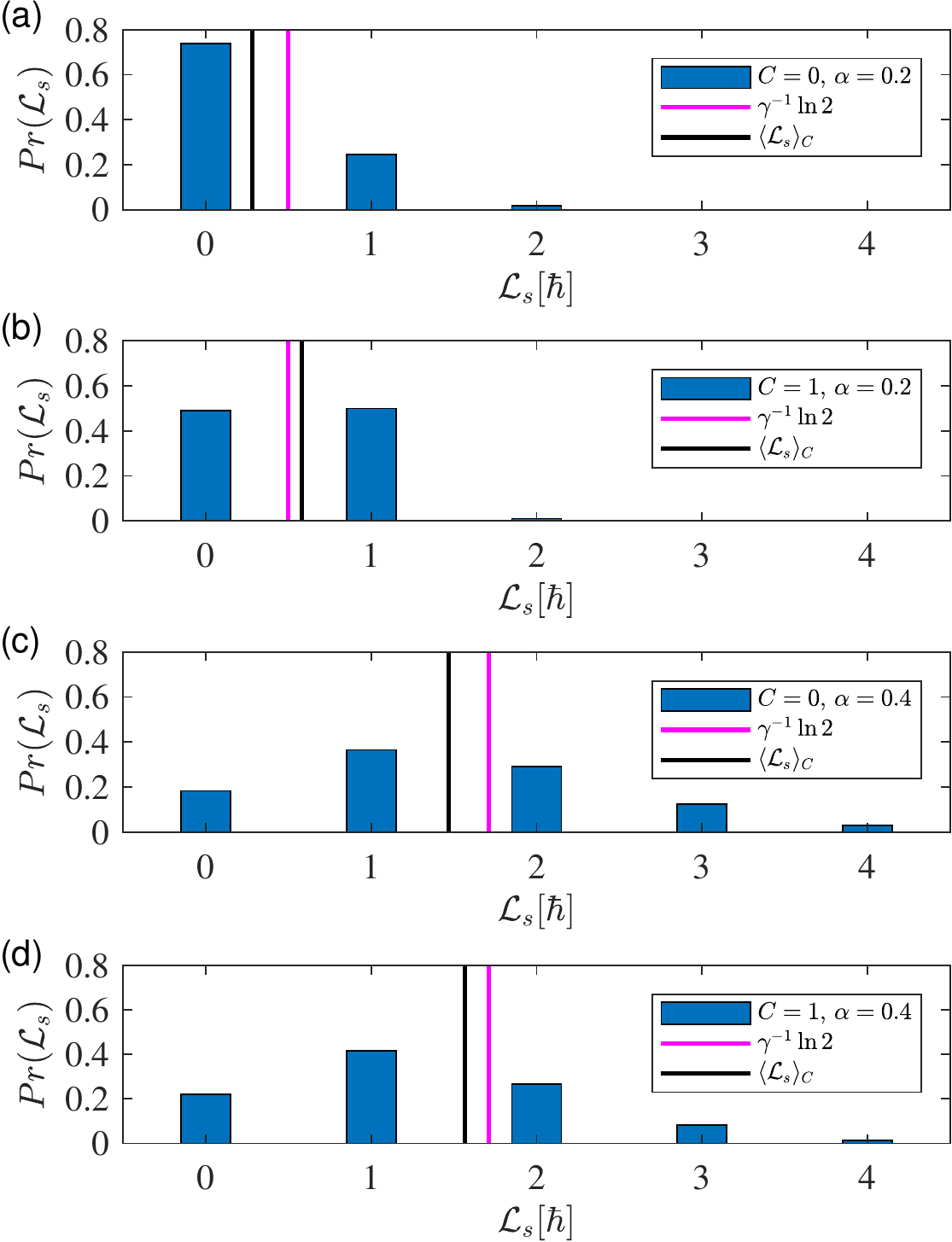}
	\caption{Spinlabor distribution for different protocols with $p=0.5$. The black line indicates the average value $\ip{\L}_C$, of the spinlabor performed on the memory-ancilla system, and the pink line indicates the bound on the erasure cost, $\gamma^{-1} \ln 2$, derived in Refs.~\cite{Vaccaro2011,Barnett2013} and quoted in \eq{eqn:VB}.  As discussed in the main text, a careful analysis shows that the erasure cost in Refs.~\cite{Vaccaro2011,Barnett2013} is defined in terms of the spintherm absorbed by the reservoir, and panels (a), (c) and (d) demonstrate that the bound does not apply to the average spinlabor. This highlights the need for care when considering the physical form of the erasure cost associated with a spin reservoir.} 
	\label{fig:work_dist}
	\end{figure}

Fig. \ref{fig:work_dist} compares the distributions $Pr(\L)$ for protocol variations corresponding to $C=0$ and $C=1$, and two different values of the reservoir spin polarisation $\alpha=0.2$ and $\alpha=0.4$ for the maximal-stored-information case with $p_\up=p_\dn=0.5$. The black vertical lines represent the corresponding average spinlabor cost $\ip{\L}_C$ calculated using Eq. \eqref{eqn:avg}, and the pink vertical lines represent the bound on the overall cost of erasure, $\gamma^{-1}\ln{2}$ in \eq{eqn:VB},  derived in Refs.~\cite{Vaccaro2011,Barnett2013}.
Notice that the distribution is rather Gaussian-like for $\alpha=0.4$; in fact, we show in \hyperref[sec:ap gaussian distribution]{Appendix \ref{sec:ap gaussian distribution}} that the distribution approaches a Gaussian distribution as $\alpha$ tends to $0.5$. 

The changing nature of the spinlabor cost distribution for different values of $\alpha$ can be traced to the relative smoothness of the spin reservoir distribution on the scale of the discreteness of the spin angular momentum spectrum during the equilibration process. The smoothness is measured by the ratio of the probabilities being sampled by the initial memory gap of $(C+1) \hbar$ of spin angular momentum for the first equilibration step,  which by \eq{eqn:probr} is given by $P_\up(C+n+1)/P_\up(n) = e^{-\gamma(C+1)\hbar}$. A vanishingly small ratio corresponds to a spin reservoir distribution that has relatively large jumps in value for consecutive spin angular momentum eigenvalues. Alternatively, a ratio that is approximately unity corresponds to a relatively smooth distribution that is amenable to being approximated as a Gaussian function as discussed in \hyperref[sec:ap gaussian distribution]{Appendix \ref{sec:ap gaussian distribution}}. Given the exponential nature of the ratio, a suitable intermediate value is $P_\up(C+n+1)/P_\up(n) = e^{-1}$. Here critical values of the ratio are, $\gamma \gg \frac{1}{(C+1)\hbar}$, $\gamma = \frac{1}{(C+1)\hbar}$, and $\gamma \ll \frac{1}{(C+1)\hbar}$ where we associate them with a ``cold'', ``warm'', and ``hot'' spin reservoir temperature, respectively. From \eq{eqn:gamma} the associated value of $\alpha$ for warm is 
\bea
\alpha = (e^\frac{1}{(C+1)}+1)^{-1}. \label{eqn:alpha critical}
\eea
Hence for $C=0$ we have $\alpha = 0.269$ and $C=1$ we have $\alpha = 0.378$. The values of $\alpha$ for Fig.~ \ref{fig:work_dist} were chosen such that panels (a) and (b) correspond to a cold spin reservoir and panels (c) and (d) correspond to a hot spin reservoir for both $C=0$ and $C=1$. Evidently, as the value of $\alpha$ increases above $0.269$ and $0.378$, the discreteness of the spin angular momentum spectrum becomes less significant and the spinlabor cost distribution approaches a Gaussian distribution.

\begin{figure}[H]
	\centering
	\includegraphics[width=0.48\textwidth]{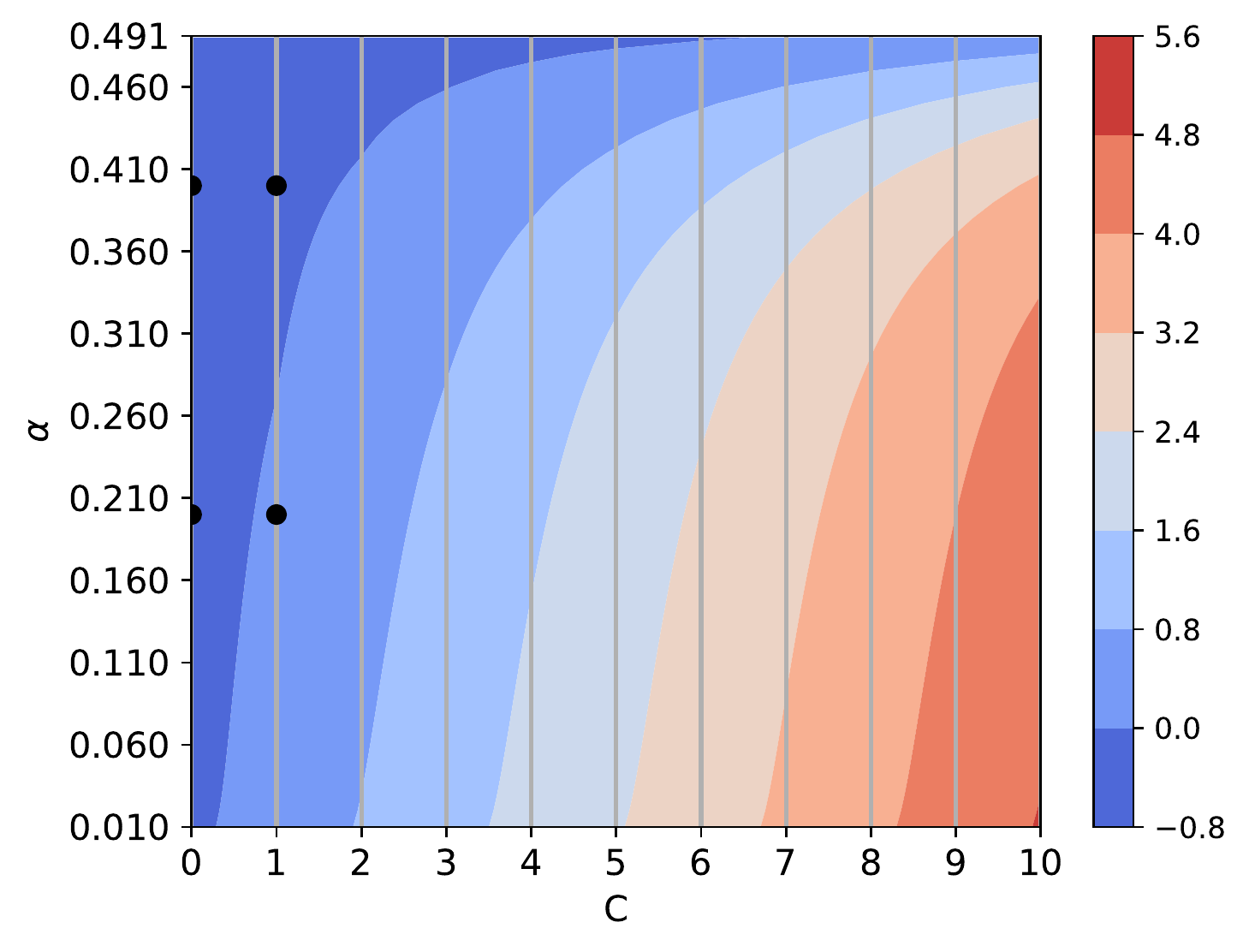}
	\caption{The values of $R$ in {\eq{eqn:R}} as a function of $C$ and $\alpha$ for the maximal-stored information case $p_\dn=p_\up=0.5$. The value of the average spinlabor cost $\ip{\L}_C$ is calculated using {\eq{eqn:avg}}, and to enhance the graphical representation, the values of $R$ have been interpolated between the discrete values of $C$ (vertical gray lines). The black dots represent the four values chosen for Fig.~\ref{fig:work_dist}}
	\label{fig:phasespace_check_spintherm}
\end{figure}

\begin{figure}[b]
	\centering
	\includegraphics[width=0.48\textwidth]{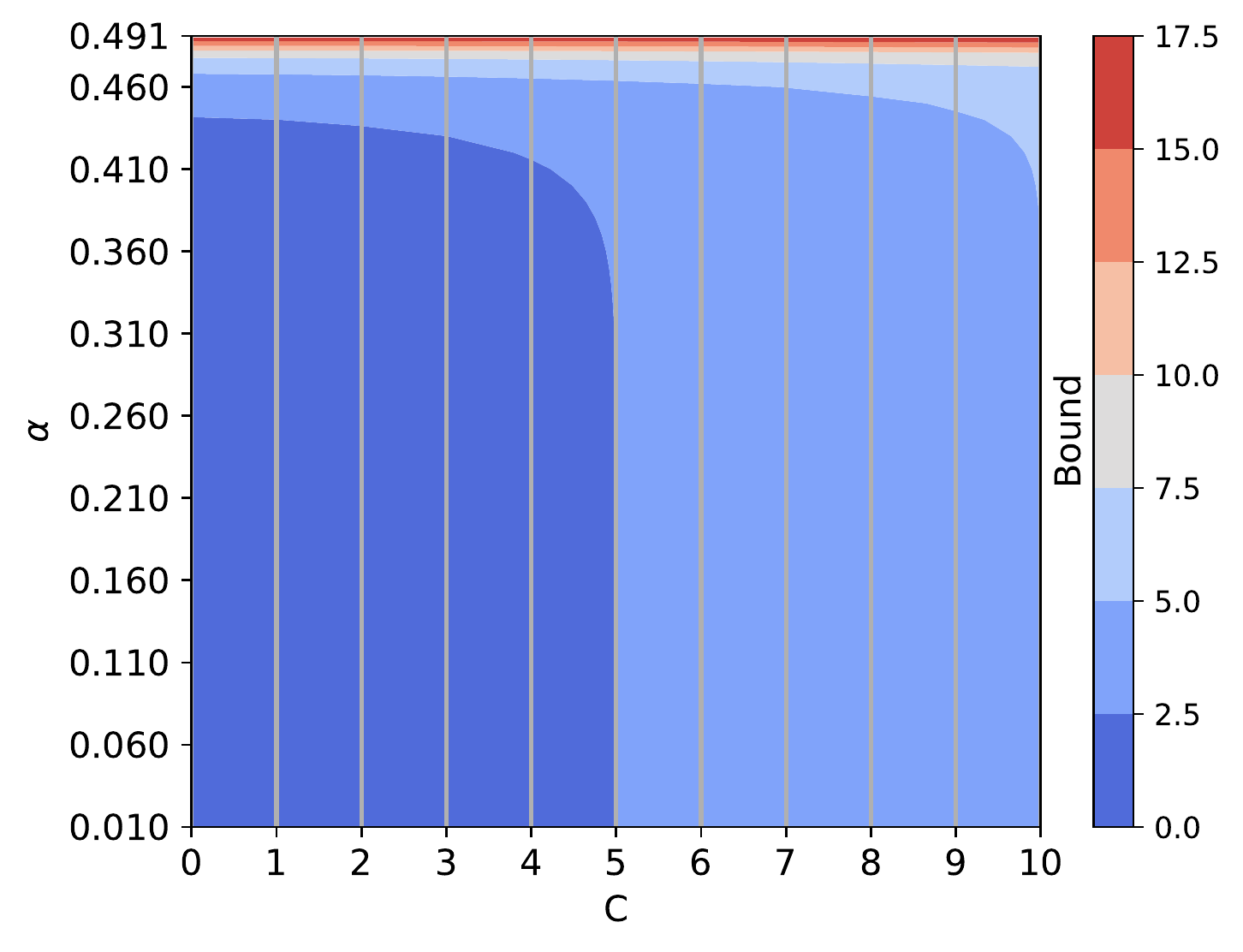}
	\caption{A contour plot, similar to Fig (3) and for the same maximal-stored information case, of the RHS of \eq{eqn:Spinlabor_ineq} as a bound on the average spinlabor cost $\ip{\L}_C$.}
	\label{fig:contour plot spinlabor}
\end{figure}

Notice that in Fig.\ref{fig:work_dist} the average spinlabor (black line) is \emph{less} than the bound (pink line) for all cases except for $C=1$ and $\alpha=0.2$. To determine why, we compare the difference 
\bea   \label{eqn:R}
    R= \ip{\L}_C-\gamma^{-1}\ln 2,
\eea
between the average accumulated spinlabor cost $\ip{\L}_C$ and the original bound as a diagnostic tool for a range of values of $C$ and $\alpha$ in Fig. \ref{fig:phasespace_check_spintherm}. 
The negative areas of the figure, shown in dark blue, are where the average spinlabor cost is less than the bound $\gamma^{-1} \ln 2$. Conversely, the other areas of the figure show a positive difference indicating that the average spinlabor cost is greater than the bound. The figure shows that for any given value of $\alpha$, the spinlabor cost increases as the value of $C$ increases, indicating that lower values of $C$ are less costly. It also shows that the increase in cost is less significant for larger values of $\alpha$, however, this is in comparison to the bound, given by $\gamma^{-1}\ln 2=\frac{\hbar\ln 2}{\ln(\alpha^{-1}-1)}$ according to \eq{eqn:gamma}, which diverges as $\alpha$ approaches 0.5.
We have collected the values of $R$ for the 4 panels in Fig. \ref{fig:work_dist} in Table. \ref{tab:R_values}. Evidently the measure $\Delta J^\prime_z$ of the cost of erasure quoted in {\eq{eqn:VB}} does not reflect the actual cost evaluated in terms of spinlabor $\ip{\L}_C$.  The reason can be traced to the derivation of {\eq{eqn:VB}} in Ref. {\cite{Vaccaro2011}} where $\Delta J^\prime_z$ is defined in Eq.~(3.9) as the spinlabor $\ip{\L}$ performed on the memory-ancilla system \emph{plus} the $\hbar/2$ of initial spintherm of the memory.  Although the spinlabor is performed on the memory-ancilla system, by the end of the erasure process it is evidently dissipated as spintherm and transferred to the reservoir under the assumed conditions of spin angular momentum conservation.  The additional $\hbar/2$ represents extra spintherm that is also evidently transferred to the reservoir under the same conditions.  As any spin angular momentum in the reservoir is in the form of spintherm, we interpret $\Delta J^\prime_z$ as the spintherm transferred to the reservoir.
\begin{table}[ht]
\centering

\begin{tabular}[t]{l|ccc}
 \hline
 Panel & $C$ & $\alpha$ & $R$ \\ 
 \hline
 a) & 0 & 0.2 & -0.22  \\ 

 b) & 1 & 0.2 & 0.08  \\

 c) & 0 & 0.4 & -0.24  \\

 d) & 1 & 0.4 & -0.14 \\
\end{tabular}
\caption{Values of R for the parameters chosen for Fig. \ref{fig:work_dist}.}
\label{tab:R_values}
\end{table}
Hence, \eq{eqn:VB} evidently bounds the erasure cost when it is expressed in terms of spintherm transferred to the reservoir---it is not specifically a bound on the spinlabor performed on the memory-ancilla system, (a more detailed analysis of the bounds are provided in \S\ref{sec:bounds}). Despite this, the bound 
serves as a basis for comparing the spinlabor cost for erasure protocols with different values of $C$, and since it was the first bound to be calculated, we shall refer to it as the \emph{original bound}.

A more direct analysis of the spinlabor cost is given by examining the expression for $\ip{\L}_C$ in \eq{eqn:avg}.
By lower-bounding the sum in Eq. \eqref{eqn:avg} with an integral using \eq{eqn:Q}, we find the bound specific for average spinlabor is given by
\bea
  \ip{\L}_C  & \geq & \hbar C p_{\uparrow}+\int_{m=C}^{\infty} \hbar Q_\uparrow(m)dm \nonumber \\
   & = & \hbar C p_{\uparrow}+\gamma^{-1} \ln(1+e^{-(C+1)\gamma \hbar}).
    \label{eqn:Spinlabor_ineq}
\eea
In Fig.~\ref{fig:contour plot spinlabor} we plot the right side of \eq{eqn:Spinlabor_ineq} as a function of $C$ and $\alpha$ for the maximal-stored information case $p_\dn=p_\up=0.5$.  The spinlabor cost clearly increases with $\alpha$, as expected, and we again find that it increases with $C$.  It is more cost efficient to delay the first CNOT step until the first equilibration step has been done, i.e. for $C=0$, for which the first term vanishes and the bound becomes $\gamma^{-1}\ln(1+e^{-\gamma \hbar})$. In this particular case the bound is lower than the original bound of $\gamma^{-1} \ln2$. Notice that $\gamma^{-1}\ln(1+e^{-\gamma \hbar})\to\gamma^{-1} \ln2$ as $\hbar\to 0$.  Thus, as $\hbar$ is the step in the discrete-valued spinlabor cost due to individual CNOT steps, we find that the difference vanishes in the continuum limit. The spin-based erasure process then becomes equivalent to the energy-based erasure processes that Landauer studied with $\gamma$ being equivalent to the inverse temperature $\beta$.

To appreciate why the $C=0$ protocol is the most efficient we need to address a subtle issue in information erasure.  Associating information erasure simply with the reduction in entropy of the memory-ancilla system carries with it the problem that erasure would then only occur, strictly speaking, during the equilibration step and the role played by the CNOT step and its associated spinlabor cost would be ignored. A better approach is to recognise that there are two types of information erasure, \textit{passive erasure} and \textit{active erasure}.
We define passive erasure as erasure that occurs without any work or spinlabor being performed and, conversely, we define active erasure as erasure that involves work or spinlabor being applied to the system. From these general definitions we can state that passive erasure takes place in the erasure protocols discussed in this section when the memory-ancilla entropy is reduced in an equilibration step \emph{without} a CNOT step preceding it. Conversely, we can state that active erasure takes place when the memory-ancilla entropy is reduced in an equilibration step \emph{with one or more} CNOT steps preceding it.
These definitions are beneficial in helping to determine if there are heat/spintherm or work/spinlabor cost occurring within a protocol Ref \cite{Briones2017}. For example, the authors of Ref \cite{Clivaz2019} make this distinction when stating that to make a non-trivial change in a target system an external coherent control or rethermalization with a thermal bath must be applied to the target which for our case the target system is the memory being erased.

The distinction between the two types of erasure is evident in the difference between erasure protocols with $C=0$ and $C=1$. In the case of $C=0$, there is no CNOT step preceding the first equilibration step, and so the reduction in entropy it produces is an example of passive erasure. Thereafter, every equilibration step is preceded by a CNOT step and so the remainder of the protocol consists of active erasure. In contrast, the case of $C=1$ entails a CNOT step before every equilibration step, including the first, and so the protocol consists of entirely of active erasure.  The important points here are that only active erasure is associated with a spinlabor cost, and the active erasure parts of both protocols are operationally identical. It then becomes clear why the protocol for $C=0$ incurs the lower spinlabor cost: it takes advantage of spinlabor-free passive erasure to reduce the entropy of the memory system first, before following the same spinlabor-incurring active erasure protocol as the protocol for $C=1$ but with an easier task due to the lower entropy of the memory .

The situation is rather different when we examine the spintherm cost of information erasure, as we do in the following subsection, because spintherm is transferred from the memory-ancilla system to the spin reservoir in both passive and active erasure.

\subsection{First law and spintherm cost \label{sec:first law and spintherm}}

In contrast to the spinlabor, which is applied directly to the memory-ancilla system, the spintherm cost of the erasure process is the amount of spintherm transferred from the memory-ancilla system to the spin reservoir. It is regarded as a cost because it reduces the spin polarization of the reservoir and thus, in principle, it reduces the ability of the reservoir to act as an entropy sink for future erasure processes.

During a CNOT step, the change in spin angular momentum of the memory-ancilla system is given by \eq{eqn:first_law} with $\Q=0$ as there is no transfer of spintherm from it, and so $\DJz^\M =\L$.  Here and below, we use a superscript $\M$, $\R$ or $\T$ to label the spin angular momentum $J_z$ of the memory-ancilla, reservoir or combined memory-ancilla-reservoir system, respectively.
During the equilibration step, the memory exchanges spintherm only and there is no spinlabor cost, hence $\DJz^\M= \Q$ and $\L=0$ from the first law \eq{eqn:first_law} applied to the memory-ancilla system. The corresponding changes to the reservoir are given by $\DJz^\R=0$ during a CNOT step and  $\DJz^\R=-\Q$ during an equilibration step, and so the changes to the combined system are given by
\begin{align}
    \DJz^\T &= \left\{\ary{l}{\L,\text{ during a CNOT step,}\\
                             0,\text{ during an equilibrium step}
                    }\right.
\end{align}
where $\DJz^\T=\DJz^\M+\DJz^\R$.  This is the description of the erasure process in terms of the first law for the conservation of spin angular momentum.

We use \eq{eqn:Q_s} to calculate the accumulated spintherm cost as follows.  As the first equilibration step occurs after $C$ CNOT steps, the value of $g(m_j)$ is equal to $C+1$ because the equilibration between the memory-ancilla system and the reservoir involves the exchange of spin angular momentum in multiples of $(C+1)\hbar$, and the value of $\Delta p(j,m_j)$, which is the change in the probability of the memory-ancilla system being in the spin-up state, is $Q_\up(C)-p_\up$. The spintherm cost for the first equilibration step is therefore given by

\bea
     \ip{\Q}_{C,C} = (C+1)\hbar  \left[Q_\up(C)-p_{\uparrow}\right]        \label{eqn:mlc}
\eea
where the symbol $\ip{\Q}_{C,m}$ represents the average spintherm associated with the equilibration step that occurs after the $m$-th CNOT step, and $C$ indicates the protocol variation.
For the second equilibration step $g(m_j)=C+2$,  $\Delta p(j,m_j)=Q_\up(C+1)-Q_\up(C)$, $m=C+1$, and so
\bea
    \ip{\Q}_{C,C+1} &=& (C+2) \hbar  \left[Q_\up(C+1)-Q_\up(C)\right] .
\eea
In general, it follows that for $m>C$
\bea
    \ip{\Q}_{C,m} &=& (m+1)\hbar \left[Q_\up(m)-Q_\up(m-1) \right]. \label{eqn:mthspinth}
\eea

The spintherm is additive and so taking the sum of $\ip{\Q}_{C,m}$ over $m$ from $m=C$ to infinity gives with the accumulated spintherm cost $\ip{\Q}_C$ for the entire erasure process, i.e.
\begin{align}
    \ip{\Q}_C&=\sum_{m=C}^\infty \ip{\Q}_{C,m}\notag\\
     &= (C+1)\hbar  \left[Q_\up(C)-p_{\uparrow}\right]\notag\\
     &\quad+\sum_{m=C+1}^{\infty}(m+1)\hbar \left[Q_\up(m)-Q_\up(m-1)\right]\notag\\
     &= -\sum_{m=C}^{\infty}\hbar Q_\up(m)-(C+1)\hbar p_{\up}\notag\\
     &=-\ip{\L}_C-\hbar p_{\up} \label{eqn:total spintherm cost}
\end{align}
where we have used \eq{eqn:avg} in the last line.
As expected, the accumulated spintherm $\ip{\Q}_C$ in \eq{eqn:total spintherm cost} is negative since spintherm is being transferred from the memory to the reservoir. It is interesting to note that the total spintherm cost is simply the average spinlabor cost plus an additional $ \hbar p_{\uparrow} $. Evidently, all the spinlabor applied to the memory-ancilla system during the CNOT steps is dissipated as spintherm as it is transferred, along with the spintherm of $\hbar p_{\up}$ associated with the initial entropy of the memory, to the reservoir during the equilibration steps. We can immediately write down the bound for the total spintherm cost using Eq.~\eqref{eqn:Spinlabor_ineq} with \eq{eqn:total spintherm cost} as
\bea
   \ip{\Q}_C &\geq& (C+1) \hbar p_{\uparrow}+\gamma^{-1} \ln(1+e^{-(C+1)\gamma \hbar}). \label{eqn:Heat_ineq}
\eea

\section{Jarzynski-like equality}\label{sec:Jarzynski-like equality}

In this section we derive a Jarzynski equality \cite{Crooks,Sagawa2010,Jarzynski1997a,Jarzynski2004} for the erasure process, but before we do, we need to re-examine the probability distributions describing the reservoir and memory-ancilla systems in terms of phase space variables and Liouville's theorem.

\subsection{Phase space and Liouville's theorem}

In order to determine the changes in the systems, we need to express the probability distribution as a function of phase space and internal (spin) coordinates at various times during the erasure protocol.
Accordingly, let a point in phase space at the time labelled by $\lambda$ be described by the vector $\mathbf{z}^\T_\lambda \equiv(\mathbf{z}^\R_\lambda,\mathbf{z}^\M_\lambda)$ where $\mathbf{z}^\R_\lambda$ and $\mathbf{z}^\M_\lambda$ represents coordinates in the reservoir and the memory-ancilla subspaces, respectively. In particular, $\lambda=\inl$ and $\lambda=\fnl$ label the initial and final coordinates, respectively, for any given period during the erasure procedure.

Although the phase space of the memory-ancilla and reservoir systems includes both the internal spin angular momentum and external spatial degrees of freedom, the spatial degree of freedom has no effect on the erasure process due to the energy degeneracy previously discussed, and so we leave it as implied.
Thus, let the coordinate $\mathbf{z}^\R_\lambda \equiv(n_{\lambda},\nu_{\lambda})$ represents the state of the reservoir of $N$ spin-$\frac{1}{2}$ particles in which $n_\lambda$ (and $N-n_\lambda$) are in the spin-up (respectively, spin-down) state, and $\nu_\lambda=1,2,\ldots,\binom{N}{n_\lambda}$ indexes a particular permutation of the particles.
The CNOT and equilibration steps are constructed to induce and maintain correlations in the memory-ancilla system.  The result is that at any time the memory-ancilla system has effectively a single binary-valued degree of freedom associated with the spin state of the memory particle.  The fact each CNOT step correlates one more ancilla particle with the spin state of the memory particle, means that the spin angular momentum of the memory-ancilla system is given by two numbers: $\n_\lambda$ which is a binary-valued free parameter that indicates the spin direction of the memory particle, and $a_\lambda$ which is an external control parameter equal to the number of completed CNOT steps and indicates the number of ancilla particles that are correlated with the memory particle.
The coordinate  representing the state of the memory-ancilla system is therefore given by $\mathbf{z}^\M_\lambda \equiv (\n_\lambda,a_\lambda)$.
Thus, the total spin angular momentum at point $\mathbf{z}^\T_\lambda$ is given by
\bea   \label{eqn:JzT}
    J_z^{(T)}(\mathbf{z}^\T_\lambda)= J^{(R)}_{z}(\mathbf{z}^\R_\lambda)
             + J^{(M)}_{z}(\mathbf{z}^\M_\lambda),
\eea
where
\begin{align}
  J^\R_{z}(\mathbf{z}^\R_\lambda) &= (n_\lambda-\frac{1}{2}N)\hbar
      \label{eqn:JzR}   \\
      J^\M_{z}(\mathbf{z}^\M_\lambda)
          &= [\n_\lambda (a_\lambda+1)-\frac{1}{2}(\N+1)]\hbar
       \label{eqn:JzM}
\end{align}
and $\N$ is the number of ancilla spin-$\frac{1}{2}$ particles.

We also need to express the phase space density in terms of a canonical Gibbs distribution, i.e. as an exponential of a scalar multiple of the conserved quantity.  In the case here, the conserved quantity is the $z$ component of spin angular momentum, and so the density is of the form
\bea
    f_{\lambda}(\mathbf{z}^\X_{\lambda}) \propto e^{-\gamma^\X_\lambda J_z^{(X)}(\mathbf{z}^\X_{\lambda})},
     \label{eqn:generic_Gibbs_dist}
\eea
where $X\in\{R, M\}$ labels the system, and $\gamma^\X_\lambda$ represents an inverse spin temperature.  The reservoir's probability distribution, given by Eq.~\eqref{eqn:probr}, is already in this form with $X=R$, $\gamma^\R_\lambda=\gamma$ and $n_{\lambda}=n$ for $n=0,1,\ldots,N$. Indeed, as previously mentioned, throughout the entire erasure process the spin temperature $\gamma^{-1}$ of the reservoir system is assumed to remain constant due to being very large in comparison to the memory system.

In contrast, the spin temperature of the memory-ancilla system changes due to both of the CNOT and equilibration steps. After the $m$-th CNOT operation has been applied, there are only two possibilities---either the memory spin and the first $m$ ancilla spins are spin up, or all spins are spin down---and, correspondingly, there are only two non-zero probabilities involved; we shall represent these probabilities as $q_{\up,\lambda}$ and $q_{\dn, \lambda}=1-q_{\up, \lambda}$, respectively.
Thus, the inverse spin temperature corresponding to the effective canonical Gibbs distribution in \eq{eqn:generic_Gibbs_dist} for the memory-ancilla system is given by
\bea
    \gamma^\M_\lambda=
   \frac{1}{a_\lambda+1}\frac{1}{\hbar}\ln\Big(\frac{q_{\dn,\lambda}}{q_{\up,\lambda}}\Big).
   \label{eqn:gamma_M}
\eea
In particular, for a single equilibration step
\begin{align}    \label{eqn:k_i and k_f equilibration}
        a_\inl=a_\fnl=m
\end{align}
whereas for a single CNOT step
\begin{align}    \label{eqn:k_i and k_f CNOT}
        a_\inl=m \quad \text{and} \quad a_\fnl=m+1
\end{align}
where $m$ is the number of CNOT steps that have been performed at the start of the step.
Before the first equilibration step is performed, the associated probabilities are fixed at $q_{\updn, \lambda}=p_\updn$ (i.e. the initial probabilities) where, for brevity,  $x_\updn=y_\updn$ implies both $x_\up=y_\up$ and $x_\dn=y_\dn$ for arbitrary variables $x$ and $y$.  For the first equilibration step the probabilities are $q_{\updn, \inl}=p_\updn$, and $q_{\updn, \fnl}=Q_\updn(C)$ whereas for any later equilibration step the probabilities are $q_{\updn, \inl}=Q_\updn(m-1)$ and $q_{\updn,\fnl}=Q_\updn(m)$ where $Q_\updn$ is given by Eq.~\eqref{eqn:Q} and $m$ is the number of prior CNOT steps.
\eq{eqn:gamma_M} is easily verified by substitution into \eq{eqn:generic_Gibbs_dist} using $X=M$ and $J_z^{(M)}$ from \eq{eqn:JzM} to show $f_\lambda\propto q_{\updn,\lambda}$.

The distribution for the combined reservoir-memory-ancilla system at time labelled $\lambda$ is thus
\bea
   f_{\lambda}(\mathbf{z}^\T_\lambda) =\frac{e^{-\gamma J^\R_{z}(\mathbf{z}^\R_\lambda)}}{Z^\R}\frac{e^{-\gamma^\M_\lambda J^\M_{z}(\mathbf{z}^\M_\lambda)}}{Z^\M_{\lambda}}
   \label{eqn:distribution_function}
\eea
where $Z^\R$ and $Z^\M_{\lambda}$ are the respective partition functions, i.e.
\begin{align}
\begin{aligned}
     \label{eqn:partition_functions}
    Z^\R &= \sum_{\mathbf{z}^\R}e^{-\gamma J^\R_{z}(\mathbf{z}^\R)}\\
    Z^\M_{\lambda} &= \sum_{\mathbf{z}^\M_\lambda}e^{-\gamma^\M_\lambda J^\M_{z}(\mathbf{z}^\M_\lambda)}.
\end{aligned}
\end{align}
The combined reservoir-memory-ancilla system is closed except for the CNOT operations when spinlabor $\mathcal{L}_{\rm s}$ is performed on the memory-ancilla system.  By the first law \eq{eqn:first_law}, therefore, the spinlabor is equal to the change in the total spin angular momentum of the combined reservoir-memory-ancilla system, i.e.
\begin{align}    \label{eqn:defining spinlabor}
       \L(\mathbf{z}_\fnl,\mathbf{z}_{\inl}) = J^\T_z(\mathbf{z}_\fnl)-J^\T_z(\mathbf{z}_{\inl})
\end{align}
where $\mathbf{z}_{\inl}$ and  $\mathbf{z}_{\fnl}$ are the corresponding initial and final points of a trajectory in phase space.

In analogy with the definition of the stochastic work \cite{Funo2018}, $\L$ will be called the  \emph{stochastic spinlabor}. Moreover, there is a fixed relationship between $\mathbf{z}_{\inl}$ and  $\mathbf{z}_{\fnl}$ because the CNOT operation is deterministic and the combined system is closed during the equilibrium step.  The evolution of the combined reservoir-memory-ancilla system is, therefore, deterministic overall.
For the sake of brevity, we have been focusing explicitly on the internal spin degrees of freedom, however, as the deterministic description appears only when all degrees of freedom are appropriately accounted for, we assume that the coordinates $\mathbf{z}^{(\rm implicit)}_\lambda$ associated with any additional ones are included in the definition of the phase space points through an implicit extension of the kind $\mathbf{z}_\lambda\mapsto (\mathbf{z}_\lambda,\mathbf{z}^{(\rm implicit)}_\lambda)$.
Thus, the final point is implicitly a function of the initial point, i.e.
\bea
   \mathbf{z}_{\fnl} = \mathbf{z}_{\fnl}(\mathbf{z}_{\inl}),
   \label{eqn:final_as_function_of_initial}
\eea
and dynamics of the combined reservoir-memory-ancilla system follows Liouville's theorem \cite{Jarzynski1997a,Jarzynski1997b} in the following form
\bea
     f_\fnl(\mathbf{z}_\fnl) = f_\inl(\mathbf{z}_\inl)
     \label{eqn:Liouville_eqn}
\eea
where $f_\inl(\mathbf{z})$ and $f_\fnl(\mathbf{z})$ are the initial and final probability distributions with respect to phase space variable $\mathbf{z}$.

\subsection{Jarzynski-like equality and probability of violation}

We are now ready to derive an expression that is analogous to the equality
\begin{align}    \label{eqn:Jarzynski's original equation}
        \langle e^{-\beta (W - \Delta F) }\rangle=1
\end{align}
where $\beta$ is the inverse temperature of a thermal reservoir, $W$ is the work performed on a system that is in quasiequilibrium with the reservoir, and $\Delta F$ is the change in the system's free energy, derived by Jarzynski \cite{Crooks,Sagawa2010,Jarzynski1997a,Jarzynski2004}. In contrast to the quasiequilibrium conditions associated with \eq{eqn:Jarzynski's original equation}, the spinlabor is performed in our erasure protocols while the memory-ancilla system is decoupled from the spin reservoir, and the equilibration steps---which re-establish equilibrium with the reservoir---are distinct operations.
In our previous paper \cite{Croucher2017}, we derived the Jarzynski-like equality,
\begin{align}    \label{eqn:Previous Jar-like equality}
            \langle e^{-\gamma \L+\ln 2} \rangle_1 = \frac{1+e^{-\gamma\hbar}}{1+e^{-2\gamma\hbar}},
\end{align}
for the protocol corresponding to $C=1$ with initial memory probabilities $p_\up=p_\dn=0.5$.
The fact that the right side is not unity shows that the ``exponential average'' \cite{Jarzynski1997a} of the spinlabor,
\bea    \label{eqn:exponential average definition}
    \ip{\L}^{\rm exp}\equiv -\gamma^{-1}\ln[\langle e^{-\gamma \L} \rangle],
\eea
deviates from the original bound of $\gamma^{-1}\ln 2$. We now generalise this result for arbitrary protocols. We begin by noting that the phase-space points $\mathbf{z}^\M_{\inl}$ and $\mathbf{z}^\M_{\fnl}$ occupied by the memory-ancilla system before and after any equilibration step
are statistically independent. This implies that the spinlabor performed on the memory-ancilla system before and after this step are also statistically independent.
With this in mind, we divide the total spinlabor into two parts as $\mathcal{L}_{\rm s}=\mathcal{L}^{(1)}_{\rm s}+\mathcal{L}^{(2)}_{\rm s}$ where superscripts $(1)$ and $(2)$ label the period where the spinlabor is performed as follows:
\begin{list}{(\arabic{enumii})}{\usecounter{enumii}}
	\item the period up to just prior to the first equilibration step, and
	\item the period following the first equilibration step to the end of the erasure process.
\end{list}
We omit in the intermediate period covering the first equilibration step because it incurs no spinlabor cost and so $\mathcal{L}_{\rm s}$ is identically zero.
Consider the expression $\langle e^{-\gamma\mathcal{L}_{\rm s}}\rangle_C$ containing the spinlabor scaled by the inverse spin temperature of the reservoir, factorised according to the statistical independence, as follows
\bea
    \langle e^{-\gamma\mathcal{L}_{\rm s}}\rangle_C
    &=& \langle e^{-\gamma\mathcal{L}^{(1)}_{\rm s}-\gamma\mathcal{L}^{(2)}_{\rm s}}\rangle_C\nonumber \\
    &=& \langle e^{-\gamma\mathcal{L}^{(1)}_{\rm s}}\rangle_C \langle e^{-\gamma\mathcal{L}^{(2)}_{\rm s}} \rangle_C
    \label{eqn:product of exp spinlabor}
\eea
where the subscript $C$ indicates the variation of the protocol in accord with \eq{eqn:avg}. The general form of each factor on the right side, with the spinlabor written in terms of the change in total spin angular momentum, is
\begin{align}
    \langle e^{-\gamma \mathcal{L}^\x_{\rm s}} \rangle
     = \sum_{\mathbf{z}^\T_\inl}  f_\inl(\mathbf{z}^\T_\inl) e^{-\gamma\big[ J^\T_z(\mathbf{z}^\T_\fnl)- J^\T_z(\mathbf{z}^\T_{\inl})\big]} \label{eqn:general_form}
\end{align}
where $x=1$ or $2$ labels the part of the spinlabor, $\mathbf{z}^\T_\inl$ and $\mathbf{z}^\T_\fnl$ are the initial and final points of the corresponding period where the spinlabor is performed, and  Eqs.~\eqref{eqn:final_as_function_of_initial} and \eqref{eqn:Liouville_eqn} are assumed to hold.

In the case of period (1), the possibilities for $\mathbf{z}^\M_\lambda=(\n_\lambda,a_\lambda)$ are either $\n_\inl=\n_\fnl=0$ or $\n_\inl=\n_\fnl=1$ with $a_\inl=0$ and $a_\fnl=C$, and the initial distribution given by Eq.~\eqref{eqn:distribution_function} reduces to
\bea   \label{eqn:initial_distribution}
       f_\inl(\mathbf{z}^\T_\inl) =\frac{e^{-\gamma J^\R_{z}(\mathbf{z}^\R_\inl)}}{Z^\R}
       \left\{ \begin{array}{l} p_\up, \mbox{ for }\n_{\inl}=1\\
       p_\dn, \mbox{ for } \n_{\inl}=0
       \end{array}\right\}
\eea
Using Eqs. \eqref{eqn:JzM}, \eqref{eqn:partition_functions} and \eqref{eqn:initial_distribution} then gives
\begin{align}
    \langle e^{-\gamma \mathcal{L}^{(1)}_{\rm s}} \rangle_C
    &= \Big(\sum_{\mathbf{z}^\R} \frac{e^{-\gamma J^\R_{z}(\mathbf{z}^\R)}}{Z^\R}\Big) \notag\\
    &\times\Big(p_\dn + p_\up e^{-\gamma \big\{[C+1-\frac{1}{2}(\N+1)]\hbar-[1-\frac{1}{2}(\N+1)]\hbar\big\}}    \Big) \notag\\
    &= p_\dn + p_\up e^{-\gamma C\hbar}.
    \label{eqn:spinlabor_period_1}
\end{align}
For future reference, we also find that
\begin{align}    \label{eqn:Z_i for period (1)}
        Z^\M_\inl = \frac{e^{\frac{1}{2}(\N+1)\ln \frac{p_\dn}{p_\up}}}{p_\dn}
\end{align}
from Eq \eqref{eqn:partition_functions}.

Period (2) begins immediately after the first equilibration step when the $\M$ system has the same spin temperature as the reservoir. Substituting for $f_\inl(\mathbf{z}^\T_\inl)$ in Eq.~\eqref{eqn:general_form} using Eqs.~\eqref{eqn:distribution_function} and \eqref{eqn:partition_functions} with $\gamma^\M_\inl=\gamma$, setting $x=2$ and again using Eq.~\eqref{eqn:final_as_function_of_initial} gives
\begin{align}
    \langle e^{-\gamma \mathcal{L}^{(2)}_{\rm s}} \rangle_C
    &= \sum_{\mathbf{z}^\T_\inl}
        \frac{e^{-\gamma J^\T_{z}(\mathbf{z}^\T_\inl)}}{Z^\R Z^\M_\inl} e^{-\gamma \big[J^\T_z(\mathbf{z}^\T_\fnl)-J^\T_z(\mathbf{z}^\T_{\inl})\big]} \notag\\
    &=\frac{Z^{\M}_\fnl}{Z^\M_\inl}.  \label{eqn:spinlabor_period_2a}
\end{align}
The possibilities for $\mathbf{z}^\M_\inl=(\n_\inl,a_\inl)$ here are $\n_\inl=0$ or $1$ with $a_\inl=C$,
and the corresponding values of $J^\M_{z}(\mathbf{z}^\M_\inl)$ using Eq.~\eqref{eqn:JzM} are $-\frac{1}{2}(\N+1)\hbar$ and $[C+1-\frac{1}{2}(\N+1)]\hbar$, and so from Eq.~\eqref{eqn:partition_functions} we find $Z^\M_\inl=e^{\frac{1}{2}(\N+1)\gamma\hbar}(1+e^{-(C+1)\gamma\hbar})$.
The maximum number of CNOT steps that can be performed is equal to the number of ancilla particles $\N$, i.e. $m=\N$ and so $a_\fnl=\N$. In this maximal case, the memory is the closest it can be brought to a completely erased state, for which the residual probability of the spin-up state is $Q_\up(\N)=e^{-(\N+1)\gamma\hbar)}/[1+e^{-(\N+1)\gamma\hbar)}]$ from \eq{eqn:Q}, and the ancilla particles approach their initial states.
In particular, the values of $\n_\fnl$ in $\mathbf{z}^\M_\fnl=(\n_\fnl,a_\fnl)$ are $\n_\fnl=0$ and $1$ with probabilities $Q_\dn(\N)=1-Q_\up(\N)$ and $Q_\up(\N)$, respectively, and as
\begin{align}    \label{eqn:JzM(z_f)}
     J^\M_{z}(\mathbf{z}^\M_\fnl)=(\n_\fnl-\frac{1}{2})(\N+1)\hbar
\end{align}
from Eq.~\eqref{eqn:JzM}, the corresponding value of the partition function in \eq{eqn:partition_functions} is $Z^\M_\fnl = e^{\frac{1}{2}(\N+1)\gamma\hbar} +e^{-\frac{1}{2}(\N+1)\gamma\hbar}$.
In the limit that the number of ancilla spins is large, i.e. $\N\gg 1$, \footnote{We assume that the number of spins in the reservoir, $N$, is at least one larger than the number of ancilla spins $\N$. This is required to enable the equilibration step to take place, which involves the exchange of $(m+1)\hbar$ of spin angular momentum between the reservoir and the memory-ancilla system, where $m$ is the number of CNOT steps that have been performed.} we find
\begin{align}
      Z^\M_\fnl=e^{\frac{1}{2}(\N+1)\gamma\hbar}   \label{eqn:Z_f for period (2)},
\end{align}
where we have ignored the exponentially-insignificant term $e^{-\frac{1}{2}(\N+1)\gamma\hbar}$.
Hence, the limiting value of Eq.~\eqref{eqn:spinlabor_period_2a} is
\begin{align}
    \langle e^{-\gamma \mathcal{L}^{(2)}_{\rm s}} \rangle_C
    &=\frac{1}{1+e^{-(C+1)\gamma\hbar}}.  \label{eqn:spinlabor_period_2b}
\end{align}
Substituting results Eqs.~\eqref{eqn:spinlabor_period_1} and \eqref{eqn:spinlabor_period_2b} into Eq.~\eqref{eqn:product of exp spinlabor} and setting $p_\up=p_\dn=0.5$ we find
\bea
  \langle e^{-\gamma\mathcal{L}_{\rm s}}\rangle_C
      &=&\frac{A}{2} \label{eqn:Jarzynski_2}
\eea
where we have defined
\bea
    A &\equiv&  \frac{1+e^{-C\gamma\hbar}}{1+e^{-(C+1)\gamma\hbar}}
                   \label{eqn:Jarzynski_2 A}
\eea
in agreement with our previous result \eq{eqn:Previous Jar-like equality} for $C=1$. We refer to this as our \emph{Jarzynski-like} equality for information erasure using a spin reservoir.

In analogy with the definition of the free energy, we define the \emph{free spin angular momentum} as
\begin{align}    \label{eqn:free spin angular momentum}  \mathcal{F}_{\rm s}\equiv-\gamma^{-1}\ln(Z),
\end{align}
and so its change over the times labelled $\inl$ and $\fnl$ for the memory-ancilla system is
\begin{align}    \label{eqn:change in F_s}
     \Delta \mathcal{F}_{\rm s}= -\gamma^{-1} \ln \frac{Z^\M_\fnl}{Z^\M_\inl}.
\end{align}
Accordingly, we find from \eq{eqn:spinlabor_period_2a} that $\langle e^{-\gamma \mathcal{L}^{(2)}_{\rm s}} \rangle_C= e^{-\gamma \Delta \mathcal{F}_{\rm s}^{(2)}}$, which can be rearranged as
\begin{align}    \label{eq:period 2 exp av L_s = exp Delta F}
   \langle e^{-\gamma (\mathcal{L}^{(2)}_{\rm s}-\Delta \mathcal{F}_{\rm s}^{(2)})} \rangle_C= 1
\end{align}
where $\Delta \mathcal{F}_{\rm s}^{(2)}$ is the change in memory-ancilla free spin angular momentum for period (2).  \eq{eq:period 2 exp av L_s = exp Delta F} is in the same form as Jarzynski's original result, \eq{eqn:Jarzynski's original equation}, as expected for spinlabor performed on the memory-ancilla system while it is in stepwise equilibrium with the reservoir. This is not the case for period (1) where the spinlabor is performed before the first equilibration step.

We calculate the change $\Delta \mathcal{F}_{\rm s}=-\gamma^{-1}\ln(Z^\M_\fnl/Z^\M_\inl)$ for the entire erasure process using $Z^\M_\inl$ for period (1), \eq{eqn:Z_i for period (1)}, and $Z^\M_\fnl$ for period (2), \eq{eqn:Z_f for period (2)}, to be
\begin{align}
        \Delta \mathcal{F}_{\rm s}
        &=-\gamma^{-1}\Big[\frac{1}{2}(\N+1)\Big(\gamma\hbar-\ln \frac{p_\dn}{p_\up}\Big)+\ln p_\dn \Big] \\
                     \label{eqn:Delta F for entire erasure}
        &=-\gamma^{-1}\Big[\frac{1}{2}(\N+1)\hbar\Big(\gamma-\gamma_{\inl}^{(M)}\Big)+\ln p_\dn \Big],
\end{align}
where in the last expression $\gamma_{\inl}^{(M)}$ is the initial inverse spin temperature of the memory-ancilla system at the start of the erasure procedure, and is given by \eq{eqn:gamma_M} with $a_{\inl}=0$.
Thus, we find using \eq{eqn:Jarzynski_2} and \eq{eqn:Delta F for entire erasure} that
\begin{align}    \label{eqn:Jarzynski-like with free spin}
      \langle e^{-\gamma (\mathcal{L}_{\rm s}-\Delta \mathcal{F}_{\rm s})} \rangle_C
      &= \frac{A}{2} e^{\gamma \Delta \mathcal{F}_{\rm s}}=A e^{-\frac{1}{2}(\N+1)\gamma\hbar}
\end{align}
and so
\begin{align}    \label{eqn:Jarzynski-like compared to free spin}
      \langle e^{-\gamma \mathcal{L}_{\rm s}} \rangle_C
      &=A e^{-\frac{1}{2}(\N+1)\gamma\hbar} e^{-\gamma \Delta \mathcal{F}_{\rm s}}
\end{align}
where we have set $p_\up=p_\dn=0.5$.
\eq{eqn:Jarzynski-like with free spin} generalizes our previous result given in \eq{eqn:Previous Jar-like equality}.
\eq{eqn:Jarzynski-like compared to free spin} shows that the exponential average \cite{Jarzynski1997a} of the spinlabor, $\ip{\L}^{\rm exp}_C\equiv -\gamma^{-1}\ln[\langle e^{-\gamma \L} \rangle_C]$, overestimates the change in free spin angular momentum $\Delta \mathcal{F}_{\rm s}^\M$ by $-\gamma^{-1}\ln A+\frac{1}{2}(\N+1)\hbar$. The least overestimation occurs for $C=0$ which corresponds, according to \eq{eqn:Spinlabor_ineq}, to the most efficient erasure protocol. The only way for the exponential average of the spinlabor to estimate the change in free spin angular momentum \emph{exactly}, i.e. for
\bea   \label{eqn:L_a^exp = Delta F}
  \ip{\L}^{\rm exp}_0 = \Delta \mathcal{F}_{\rm s},
\eea
is if the memory particle is in equilibrium with the reservoir at the start of the erasure procedure, in which case $p_\up=Q_\up(0)$ and $p_\dn=1-p_\up$ where $ Q_\up(m)$ is given by \eq{eqn:Q}.

Applying Jensen's inequality $\ip{f(X)}\ge f(\ip{X})$ for convex function $f$ and random variable $X$ \cite{Jensen1906} to Eq. \eqref{eqn:Jarzynski_2} yields a new lower bound on the spinlabor cost,
\bea
  \langle  \mathcal{L}_{\rm s}\rangle_C  \geq \gamma^{-1}\ln \frac{2}{A} \label{eqn:Bound 2b}
\eea
as an alternative to the bound we derived in \eq{eqn:Spinlabor_ineq}---we defer comparing these bounds until \S\ref{sec:bounds}. Also, applying Jarzynski's argument, in relation to the inequality
$e^{-X_0}\int_{-\infty}^{X_0}P(X)dX\le \int_{-\infty}^{\infty}e^{-X}P(X)dX$ for probability distribution $P(X)$ \cite{Jarzynski1999}, to Eq. \eqref{eqn:Jarzynski_2} gives the probability of violation as
\bea
      Pr^{(v)}(\epsilon) \leq e^{-\gamma \epsilon}. \label{eqn:bound A2}
\eea
Here $Pr^{(v)}(\epsilon)$ is the probability that the spinlabor cost $\mathcal{L}_{\rm s}$ violates the bound $\gamma^{-1} \ln 2 / A$ by $\epsilon$ or more (i.e the probability that $\mathcal{L}_{\rm s} \leq \gamma^{-1} \ln 2 / A- \epsilon$).

\begin{figure}[h]
	\centering
	\includegraphics[width=0.48\textwidth]{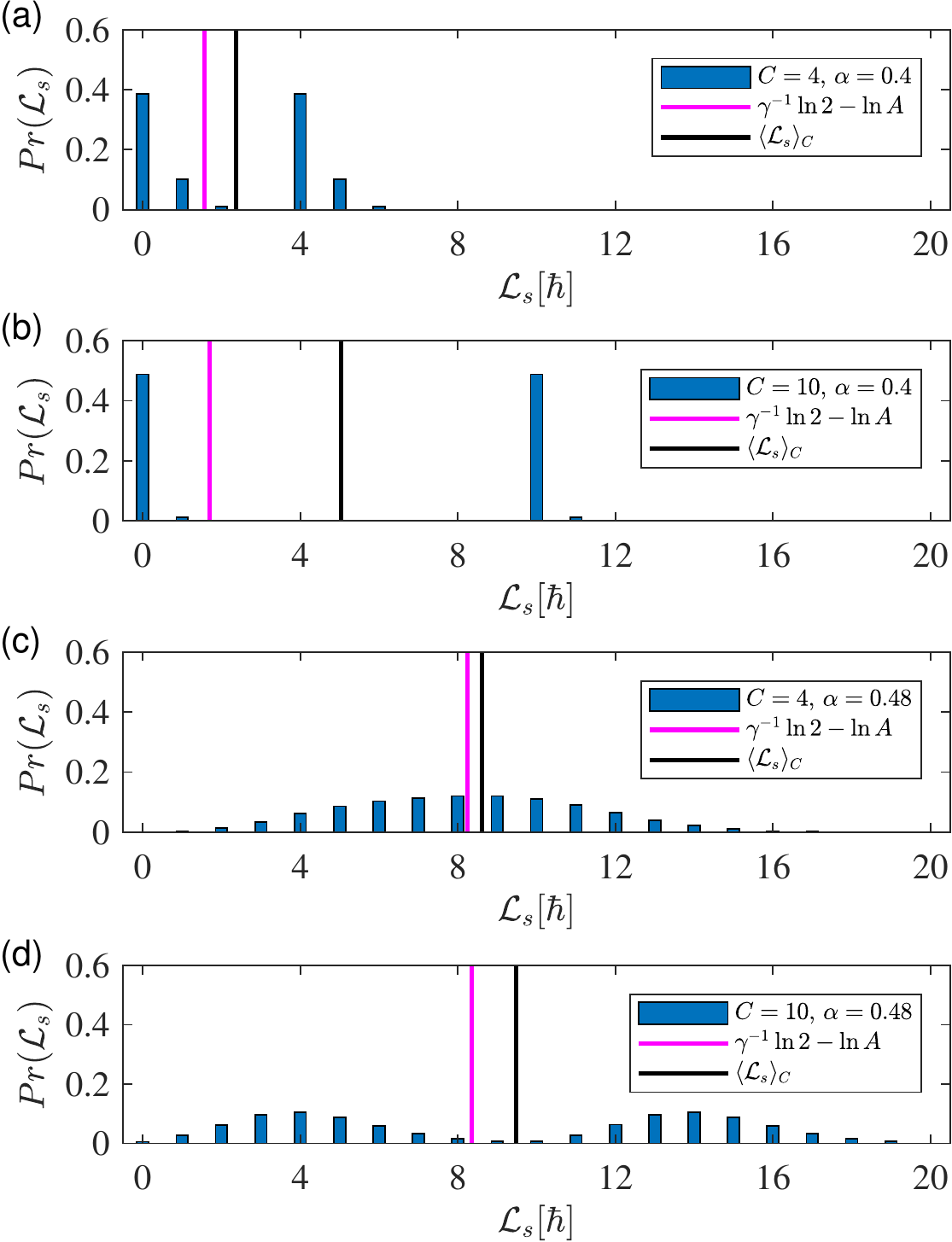}
	\caption{Spinlabor $\L$ probability distribution for different protocols for $p_{\uparrow}=0.5$. The black line indicates the average value and the pink indicates the bound of $\gamma^{-1}(\ln 2 - \ln A)$. Notice that as the average spinlabor cost in panel (c) is near the middle of the plot; this implies that there is no second peak in the distribution beyond the range plotted.
}
	\label{fig:Workdist_LargeC_sym}
\end{figure}

In Fig. \ref{fig:Workdist_LargeC_sym} we plot the spinlabor probability distributions as a function of the spinlabor $\L$ for two protocol variations, $C=4$ and $C=10$, and two reservoir spin temperatures corresponding to $\alpha=0.4$ and $\alpha=0.48$, for the maximal-stored-information case of $p_\up=p_\dn=0.5$. Applying \eq{eqn:alpha critical} for when $C=4$ and $C=10$ gives us $\alpha = 0.450$ and $\alpha = 0.478$ respectively. Hence the values of $\alpha$ were chosen to be $\alpha=0.4$ and $\alpha=0.48$ to provide us with a cold and hot distribution respectively. The distribution for when $\alpha=0.4$ is considered cold since it is less than the critical values $\alpha=0.450$ and $\alpha=0.478$ with $\gamma^{-1}=2.46$ and will have a non-gaussian like spinlabor distribution. Conversely the distribution for when $\alpha=0.48$ is considered hot since it is greater than $\alpha=0.450$ and $\alpha=0.478$ with $\gamma^{-1}=12.49$ and will have a gaussian like spinlabor distribution. Other values of $\alpha$ are not necessary since they will not provide any further information to the following analysis. The spinlabor averages (black line) are calculated using Eq. \eqref{eqn:avg} and the bound (pink line) is given by \eq{eqn:Bound 2b}.

All the averages are consistent with the bound (i.e. the black line is on the right of the pink). As previously noted in regards to Fig. \ref{fig:phasespace_check_spintherm}, we again find that the protocol becomes more expensive with increasing values of $C$. Interestingly, the distributions differ qualitatively from those in Fig.~\ref{fig:work_dist} in having two peaks separated by $\L=C$ whereas all those in Fig.~\ref{fig:work_dist} have only a single peak. The reason for the double peaks can be traced to period (1) for which the spinlabor cost depends on the initial state of the memory; that cost is either $\L^{(1)}=0$ or  $\L^{(1)}=C\hbar$ for the memory initially in the spin down and spin up states, respectively.  As the spinlabor costs incurred in periods (1) and (2) are independent and additive, the probability distributions plotted in Fig. \ref{fig:Workdist_LargeC_sym} are an average of the probability distribution describing the spinlabor cost of period (2) and a copy shifted along the $\L$ axis by $C\hbar$ which can result in a total distribution that has two separate peaks.
The exception is panel (c) for which the average spinlabor cost is in the centre of a single peak --- the spread in the spinlabor cost of period (2) is evidently of the order of the size of the shift, $C\hbar$, which results in the two peaks in the total distribution being unresolvable.
In comparison, there is no shifted copy for $C=0$ and the shift of $\hbar$ for $C=1$ does not result in a distinguishable second peak in Fig.~\ref{fig:work_dist} which is why we chose the values $C=4$ and $C=10$ for the plot and not $C=0$ or $C=1$. We also find that the distribution in the vicinity of each peak is rather Gaussian-like for $\alpha=0.48$, similar to what we found for Fig.~\ref{fig:work_dist} and demonstrated in \hyperref[sec:ap gaussian distribution]{Appendix \ref{sec:ap gaussian distribution}}.

\begin{figure}[t]
	\centering
	\includegraphics[width=0.48\textwidth]{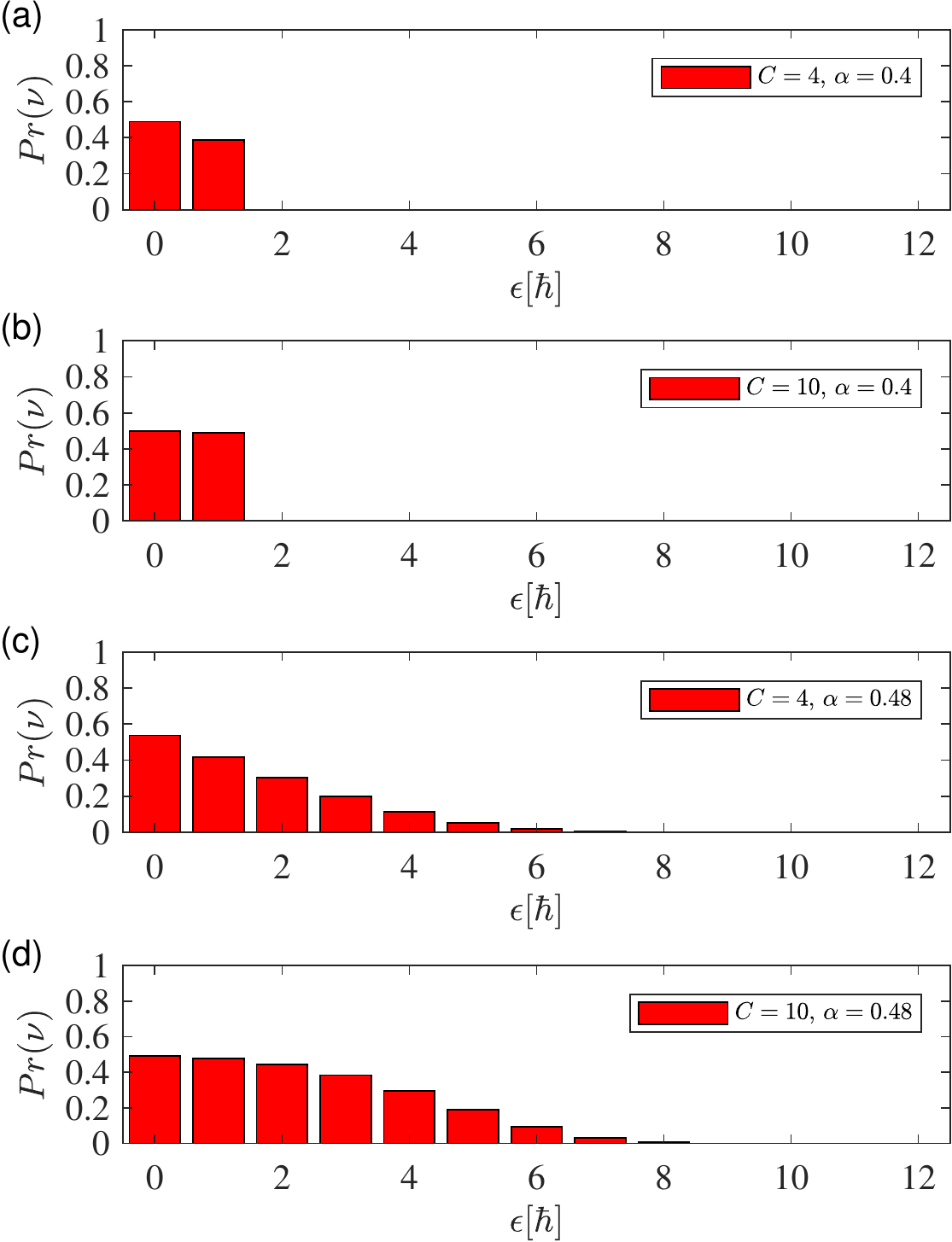}
	\caption{Probability of violation for different protocols with $p_{\up}=p_\dn=0.5$. The variables are an one
     to one correspondence the ordering in Fig. \ref{fig:Workdist_LargeC_sym}.}
	\label{fig:Probvio_LargeC_sym}
\end{figure}

In Fig. \ref{fig:Probvio_LargeC_sym} we plot the probability of violation $Pr^{(v)}(\epsilon)$ given by \eq{eqn:bound A2} as a function of $\epsilon$, for the maximal-stored-information case of $p_\up=p_\dn=0.5$. $Pr^{(v)}(\epsilon)$ is equal to the cumulative probability from $\L=0$ to $\epsilon$ below the pink line (i.e. the bound) in Fig. \ref{fig:Workdist_LargeC_sym}. We find $Pr^{(v)}(0)$ tends to $0.5$ as $C$ increases and for $\alpha$ near $0.5$ , which is not surprising given that $p_\dn=0.5$ with the figure plotting the cumulative probabilities of the left side of the pink line in Fig. \ref{fig:Workdist_LargeC_sym}.

We conclude this section with a brief analysis of the cases where the information stored in the memory is less than maximal, i.e. where $p_\dn \ne p_\up$. In these cases we find that the spinlabor bound \eq{eqn:Bound 2b} is replaced with
\bea
  \langle  \mathcal{L}_{\rm s}\rangle_C  \geq \gamma^{-1}\ln A', \label{eqn:probs not equal bound}
\eea
where
\bea
A' = \Big( \frac{p_{\downarrow} + p_{\uparrow}e^{-\gamma C \hbar}}{1+e^{-(C+1)\gamma \hbar}} \Big)
\eea
with the corresponding probability of violation, i.e. the probability that $\mathcal{L}_{\rm s} \leq \gamma^{-1} \ln A' - \epsilon$, being
\bea
      Pr^{(v)}(\epsilon) \leq e^{-\gamma \epsilon}. \label{eqn:bound A3}
\eea

\begin{figure}
	\centering
	\includegraphics[width=0.48\textwidth]{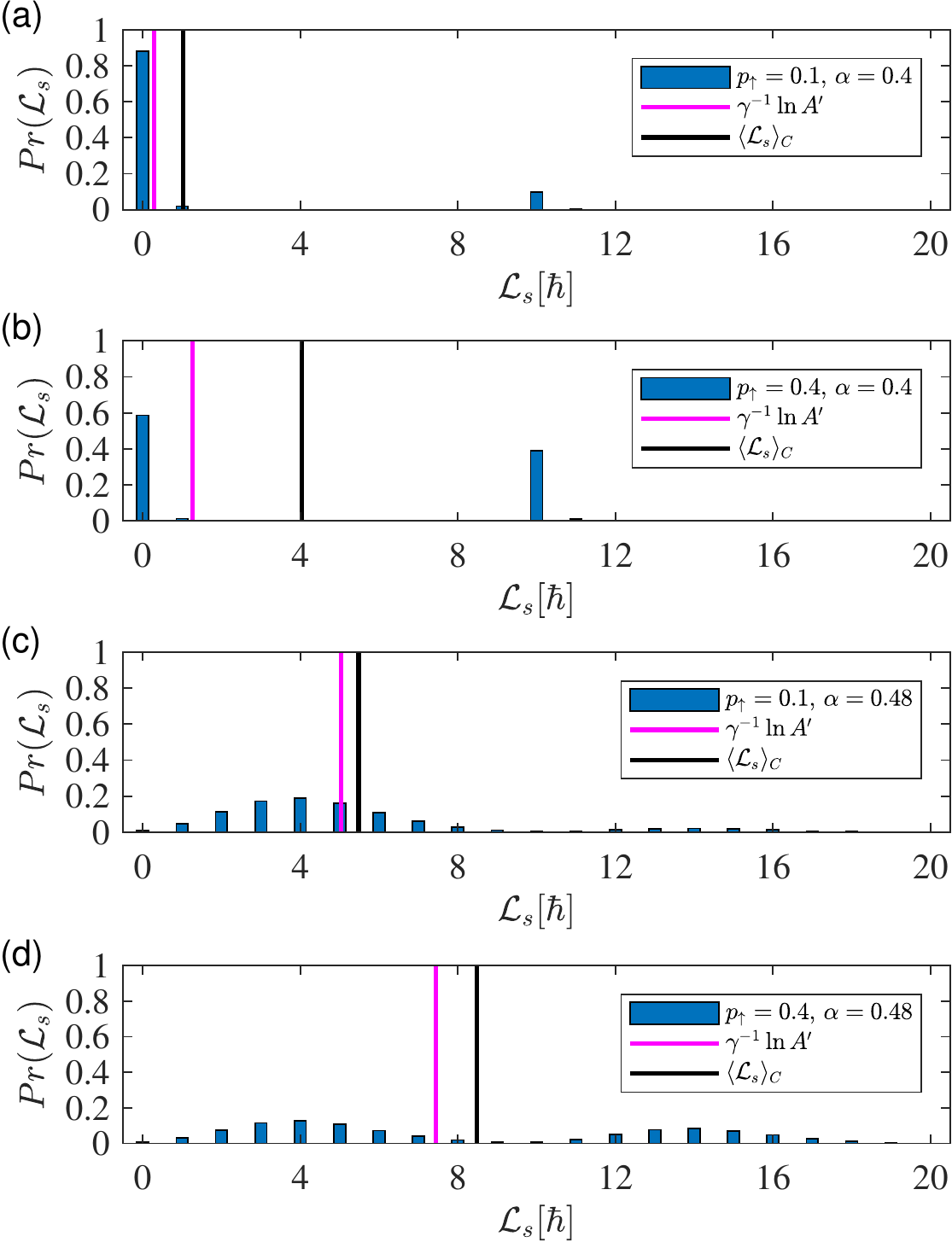}
	\caption{Spinlabor distribution for $C=10$. The black line indicates the average value and the pink $\gamma^{-1} \ln A'$.
}
	\label{fig:workasym}
\end{figure}

In Fig.~\ref{fig:workasym} we plot the spinlabor probability distributions for $p_\up=0.1$ and $p_\up=0.4$ with two different values of the reservoir spin polarization $\alpha=0.4$ and $\alpha=0.48$ for the protocol variation with $C=10$. We chose $C=10$, $\alpha=0.4$ and $\alpha=0.48$ so that these distributions can be compared directly with those in Fig. \ref{fig:Workdist_LargeC_sym}(b) and (d) for which $\alpha=0.4$ and $\alpha=0.48$, respectively, and $C=10$.  As expected from the above discussion, in each distribution in Fig.~\ref{fig:workasym} the relative height of the first peak compared to the second is found to be given by $p_\dn/p_\up$, which evaluates to $9$, $1.5$, $9$, and $1.5$ for panel (a), (b), (c) and (d), respectively; in comparison, the two peaks in each distribution in Fig.~\ref{fig:Workdist_LargeC_sym} panel (b) and (d) have the same height.

\begin{figure}[h]
\centering
\includegraphics[width=0.48\textwidth]{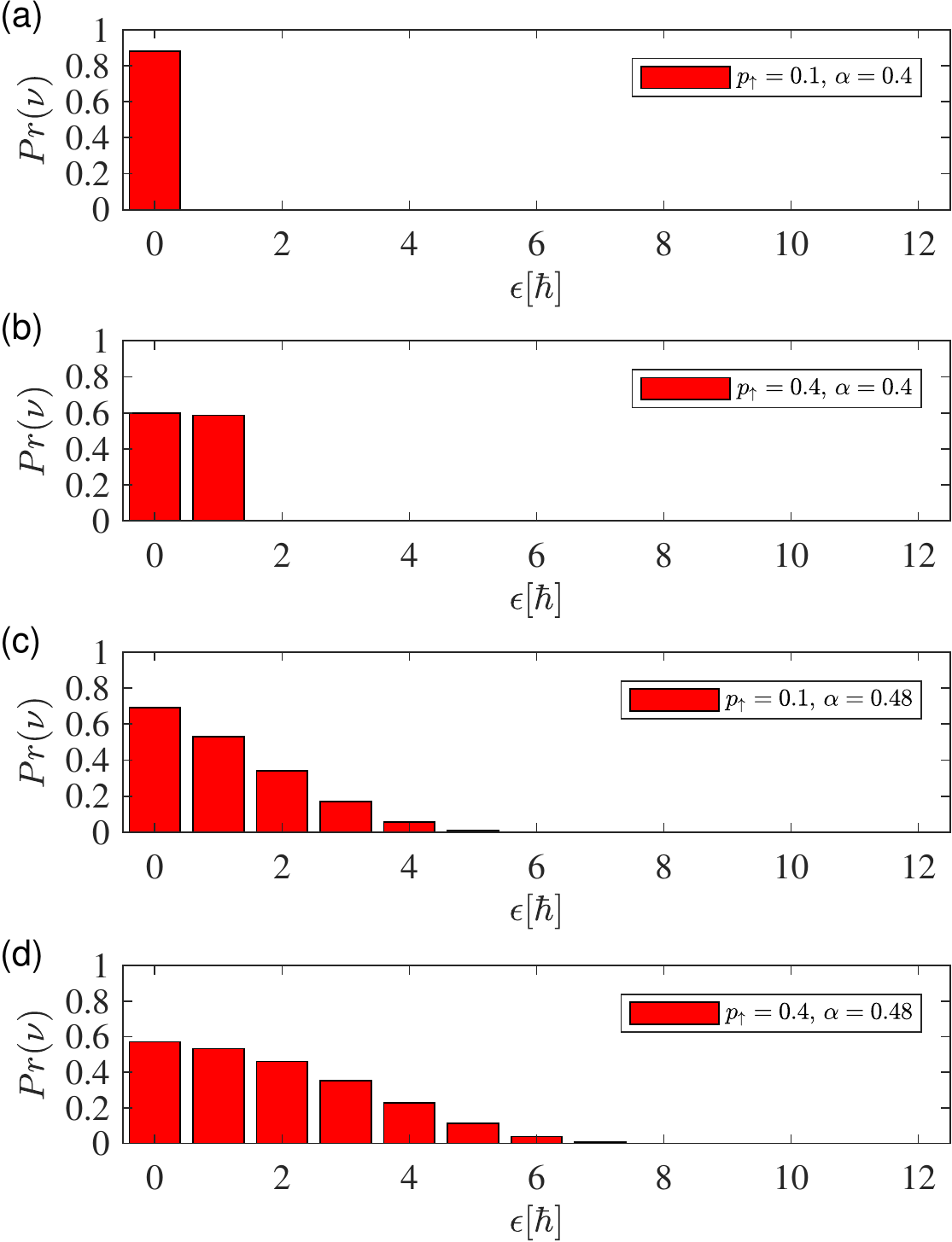}
\caption{Probability of violation for different asymmetric memory and $C=10$. The variables are a one to one correspondence to the ordering in Fig. \ref{fig:workasym}.}
  \label{fig:vioasym}
\end{figure}

The average spinlabor costs $\ip{\L}_C$ (black lines) are also lower in Fig.~\ref{fig:workasym} compared to corresponding values in Fig.~\ref{fig:Workdist_LargeC_sym} because they are associated with a higher statistical weight ($p_\dn$) for incurring the $\L^{(1)}=0$ cost. This behavior is also expected from \eq{eqn:avg} which shows that $\ip{\L}_C$ depends linearly on $p_\up$, which is correspondingly smaller.
In Fig.~\ref{fig:vioasym} we plot the probability of violation $Pr^{(v)}(\epsilon)$ for the same situations as in Fig.~\ref{fig:workasym}.  These plots are directly comparable with those in panels (b) and (d) of Fig.~\ref{fig:Probvio_LargeC_sym}. We find $Pr^{(v)}(0)$ is larger than the corresponding values in Fig.~\ref{fig:Probvio_LargeC_sym} due to the larger statistical weight (i.e. $p_\dn=0.9$ and $0.6$ in Fig.~\ref{fig:vioasym} compared to $p_\dn=0.5$ in Fig.~\ref{fig:Probvio_LargeC_sym}) of the $\L^{(1)}=0$ cost. In fact, panel (a) shows that $Pr^{(v)}(0)$ is as large as $\approx 0.9$.

\section{Integral fluctuation theorem}\label{sec:integral fluctuation theorem}
We now derive the integral fluctuation theorem for our erasure process and use it to find further bounds on the cost of spinlabor and production of spintherm. The \textit{surprisal}, also known as the stochastic Shannon entropy, associated with the probability $f(\mathbf{z})$ for the state $\mathbf{z}$ of an arbitrary system, is defined as \cite{Ince2017,Communication2012,Naghiloo,Sagawa2013}
\bea
   s(\mathbf{z})=-\ln f(\mathbf{z}). \label{eqn:Surprisal}
\eea
The average value of $s(\mathbf{z})$ is just the Shannon entropy $H= - \sum_{\mathbf{z}} f(\mathbf{z})\ln f(\mathbf{z})$. The need to introduce surprisal stems from the necessity to measure the degree of erasure for a ``single shot'' situation, such as a single cycle of the erasure protocol.
Surprisal provides more information than Shannon entropy, by allowing us to track the individual changes in information between two states in the memory as it is being erased.
The change in surprisal due to the system evolving from $\mathbf{z}_\inl$ to $\mathbf{z}_\fnl$ is given by
\cite{Deffner2011, Sagawa2012b}
\bea   \label{eqn:definition stochastic entropy production}
    \sigma(\mathbf{z}_\fnl,\mathbf{z}_\inl)=s(\mathbf{z}_\fnl)-s(\mathbf{z}_\inl) =-\ln f_\fnl(\mathbf{z}_\fnl)+\ln f_\inl(\mathbf{z}_\inl),
\eea
where $\inl$ and $\fnl$ label initial and final quantities, respectively, and is called the \emph{stochastic entropy production} of the system.

As the reservoir ($R$) and memory-ancilla system ($M$) are assumed to be statistically independent due to the relatively-large size of the reservoir, the total ($T$) stochastic entropy production of the reservoir-memory-ancilla combined system is given by the sum of the stochastic entropy production of each system, i.e. by
\begin{align}
     &\hspace{-5mm}\sigma^{(T)}(\mathbf{z}^{(T)}_\fnl,\mathbf{z}^{(T)}_\inl )= \sigma^{(R)}(\mathbf{z}^{(R)}_\fnl,\mathbf{z}^{(R)}_\inl ) +
             \sigma^{(M)}(\mathbf{z}^{(M)}_\fnl,\mathbf{z}^{(M)}_\inl )\nonumber \\
       &= -\ln f^{(R)}_\fnl(\mathbf{z}^{(R)}_\fnl)+\ln f^{(R)}_\inl(\mathbf{z}^{(R)}_\inl) \nonumber \\
       &\qquad -\ln f^{(M)}_\fnl(\mathbf{z}^{(M)}_\fnl)+\ln f^{(M)}_\inl(\mathbf{z}^{(M)}_\inl)
       \label{eqn:sigma}
\end{align}
where the probability distributions $f^{(R)}_\lambda$ and $f^{(M)}_\lambda$ are given by \eq{eqn:distribution_function}. We write the joint probability of a trajectory of the combined reservoir-memory-ancilla system that begins at $\mathbf{z}^{(T)}_\inl$ and ends at $\mathbf{z}^{(T)}_\fnl$ as
\bea
    P(\mathbf{z}^{(T)}_\fnl,\mathbf{z}^{(T)}_\inl)=P(\mathbf{z}^{(T)}_\fnl\kern-0.5em \leftarrow \mathbf{z}^{(T)}_\inl) f^{(T)}_\inl(\mathbf{z}^{(T)}_\inl) \label{eqn:joint_p}
\eea
where
\begin{align}    \label{eq:conditional probability}
   P(\mathbf{z}^{(T)}_\fnl\kern-0.5em \leftarrow \mathbf{z}^{(T)}_\inl)=\delta_{\mathbf{z}^{(T)}_\fnl(\mathbf{z}^{(T)}_\inl), \mathbf{z}^{(T)}_\inl}
\end{align}
re-expresses the deterministic trajectories relation, \eq{eqn:final_as_function_of_initial}, as the conditional probability that the total system will end at $\mathbf{z}^{(T)}_\fnl(\mathbf{z}^{(T)}_\inl)$ if it begins at $\mathbf{z}^{(T)}_\inl$.
The expression for the time reversed process is
\bea
    \tilde P(\mathbf{z}^{(T)}_\fnl,\mathbf{z}^{(T)}_\inl)=\tilde P(\mathbf{z}^{(T)}_\inl\kern-0.5em \leftarrow \mathbf{z}^{(T)}_\fnl) f^{(T)}_\fnl(\mathbf{z}^{(T)}_\fnl). \label{eqn:joint_p_tr}
\eea
The trajectories between the forward and backward processes are time symmetric, and since the combined reservoir-memory-ancilla system is either isolated from any external environment or undergoes the deterministic CNOT operation, we have
\bea
   P(\mathbf{z}^{(T)}_\fnl\kern-0.5em \leftarrow \mathbf{z}^{(T)}_\inl)
   =\tilde P(\mathbf{z}^{(T)}_\inl\kern-0.5em \leftarrow \mathbf{z}^{(T)}_\fnl).
\eea
Taking the ratio of \eqref{eqn:joint_p} and \eqref{eqn:joint_p_tr} gives
\bea
   \frac{\tilde{P}(\mathbf{z}^{(T)}_\fnl,\mathbf{z}^{(T)}_\inl)}{P(\mathbf{z}^{(T)}_\fnl,\mathbf{z}^{(T)}_\inl)}
   &=&\frac{\tilde P(\mathbf{z}^{(T)}_\inl\kern-0.5em \leftarrow \mathbf{z}^{(T)}_\fnl) f^{(T)}_\fnl(\mathbf{z}^{(T)}_\fnl)}{P(\mathbf{z}^{(T)}_\fnl\kern-0.5em \leftarrow \mathbf{z}^{(T)}_\inl) f^{(T)}_\inl(\mathbf{z}^{(T)}_\inl)} \nonumber \\
   &=&\frac{f^{(T)}_\fnl(\mathbf{z}^{(T)}_\fnl)}{f^{(T)}_\inl(\mathbf{z}^{(T)}_\inl)} ,
\eea
and then using \eq{eqn:sigma} to re-express the right side yields the detailed fluctuation theorem \cite{Alhambra2016, Sagawa2012b, Sevick2008}
\bea
   \frac{\tilde{P}(\mathbf{z}^{(T)}_\fnl,\mathbf{z}^{(T)}_\inl)}{P(\mathbf{z}^{(T)}_\fnl,\mathbf{z}^{(T)}_\inl)}
   &=& e^{-\sigma^{(T)}(\mathbf{z}^{(T)}_\fnl,\mathbf{z}^{(T)}_\inl )}  \label{eqn:DFT}
\eea
which expresses the ratio in terms of the stochastic entropy production for the erasure process.
Finally, multiplying by $P(\mathbf{z}^{(T)}_\fnl,\mathbf{z}^{(T)}_\inl)$ and summing over $\mathbf{z}^{(T)}_\inl$ and $\mathbf{z}^{(T)}_\fnl$ results in the integral fluctuation theorem \cite{Seifert2005,Seifert2012,Gong2015}
\bea  \label{eqn:integral fluctuation theorem}
    \ip{ e^{-\sigma^{(T)}} }= 1.
\eea
Using Jensen's inequality for convex functions   \cite{Jensen1906} shows that $\ip{e^{-\sigma^{(T)}}} \geq e^{-\ip{\sigma^{(T)}}}$, and so from \eq{eqn:integral fluctuation theorem} the total entropy production is
\bea  \label{eqn:total entropy production > 0}
    \ip{\sigma^{(T)}}
      \geq 0,
\eea
which expresses the non-negativity of the classical relative entropy or the Kullback–Leibler divergence $D(P(\mathbf{z}^{(T)}_\fnl,\mathbf{z}^{(T)}_\inl)||\tilde{P}(\mathbf{z}^{(T)}_\fnl,\mathbf{z}^{(T)}_\inl))$ expected from the second law \cite{Funo2018}.  This result is used below when deriving bounds on the spinlabor and spintherm costs associated with the erasure process by expressing $\sigma^\T(\mathbf{z}^{(T)}_\fnl,\mathbf{z}^{(T)}_\inl)$ in terms of either quantity.

We first focus on the spinlabor. Substituting for the probability distributions $f^{(R)}_\lambda(\mathbf{z}^{(R)}_\lambda)$ and $f^{(M)}_\lambda(\mathbf{z}^{(M)}_\lambda)$ in \eq{eqn:sigma} using the first and second factors, respectively, on the right of \eq{eqn:distribution_function} reveals
\begin{align}
      \sigma^\T(\mathbf{z}^{(T)}_\fnl,\mathbf{z}^{(T)}_\inl)
           &=\gamma J^\R_{z}(\mathbf{z}^\R_\fnl)-\gamma J^\R_{z}(\mathbf{z}^\R_\inl) \notag\\
           &\hspace{-5mm}+\gamma^{(M)}_\fnl J^\M_{z}(\mathbf{z}^\M_\fnl)-\gamma^{(M)}_\inl J^\M_{z}(\mathbf{z}^\M_\inl) \notag\\
           &\hspace{-5mm}\quad +\ln \frac{Z^\M_\fnl}{Z^\M_\inl} \label{eqn:sigma_work_pre}
\end{align}
where $\gamma$ is the constant inverse spin temperature of the reservoir, $\gamma^{(M)}_\lambda$ is the inverse spin temperature of the memory-ancilla system defined in \eq{eqn:gamma_M}, and $Z^\M_{\lambda}$ is the memory-ancilla partition function defined in \eq{eqn:partition_functions}.
There are two points to be made here.  The first is that the term for the reservoir on the right side of \eq{eqn:sigma_work_pre} corresponding to $\ln ({Z^\R_\fnl}/{Z^\R_\inl})$ is zero because the reservoir distribution $f^\R$ (and, thus, its partition function) is assumed to remain constant throughout the erasure procedure.
The second is that the inverse spin temperature of the memory-ancilla system is equal to that of the reservoir, i.e.
\begin{align}    \label{eqn:gamma_m=gamma}
      \gamma^{(M)}_\lambda=\gamma ,
\end{align}
after an equilibration step; at other times the value of $\gamma^{(M)}_\lambda$ depends on the situation as given by \eq{eqn:gamma_M}.

Recall from \eq{eqn:defining spinlabor} that the stochastic spinlabor is the change in the total spin angular momentum along a trajectory, i.e.
\begin{align}  \label{eqn:L_s(T)}
    \L(\mathbf{z}^{(T)}_\fnl,\mathbf{z}^{(T)}_\inl )
    &\equiv  J^\R_{z}(\mathbf{z}^\R_\fnl) +J^\M_{z}(\mathbf{z}^\M_\fnl)  \notag\\
    &\quad     - J^\R_{z}(\mathbf{z}^\R_\inl) - J^\M_{z}(\mathbf{z}^\M_\inl).
\end{align}
Using this, together with \eq{eqn:change in F_s}, allows us to rewrite \eq{eqn:sigma_work_pre} in terms of $\L(\mathbf{z}^{(T)}_\fnl,\mathbf{z}^{(T)}_\inl )$ and $\Delta \mathcal{F}_{\rm s}^\M$ as
\begin{align}
    \sigma^\T(\mathbf{z}^{(T)}_\fnl,\mathbf{z}^{(T)}_\inl)
     &=\gamma\Big[\L(\mathbf{z}^{(T)}_\fnl,\mathbf{z}^{(T)}_\inl ) -\Delta \mathcal{F}_{\rm s}^\M\Big]\notag \\
     &\hspace{-5mm}  + \Delta \gamma_\fnl J^\M_{z}(\mathbf{z}^\M_\fnl) - \Delta \gamma_\inl J^\M_{z}(\mathbf{z}^\M_\inl)\label{eqn:sigma_work}
\end{align}
where the last two terms account for different spin temperatures for the reservoir and memory-ancilla systems with
\begin{align}    \label{eq:Delta gamma}
      \Delta \gamma_\lambda \equiv \gamma^\M_\lambda-\gamma .
\end{align}
We are primarily interested in the initial and final states corresponding to the beginning and ending, respectively, of the entire erasure procedure where these terms are known.
In particular,  as $\mathbf{z}^\M_\inl=(\n_\inl,a_\inl)$ with $\n_\inl=0$ or $1$ with probabilities $p_\dn$ and $p_\up$, respectively, and $a_\inl=0$, we find from \eq{eqn:gamma_M} with $q_{\updownarrow,\inl}=p_\updownarrow$ that $\Delta \gamma_\inl=\frac{1}{\hbar}\ln\frac{p_\dn}{p_\up}-\gamma$, and from \eq{eqn:JzM} that
\begin{align}    \label{eqn:JzM(z_i)}
     J^\M_{z}(\mathbf{z}^\M_\inl)=[\n_\inl-\frac{1}{2}(\N+1)]\hbar.
\end{align}
For the final state, we assume that the erasure procedure ends with an equilibration step and so, according to \eq{eqn:gamma_m=gamma}, $\Delta \gamma_\fnl=0$.  Thus, for the entire erasure procedure,
\begin{align}
    \sigma^\T(\mathbf{z}^{(T)}_\fnl,\mathbf{z}^{(T)}_\inl)
     &=\gamma\Big[\L(\mathbf{z}^{(T)}_\fnl,\mathbf{z}^{(T)}_\inl ) -\Delta \mathcal{F}_{\rm s}^\M\Big]\notag\\
     &  -\Big(\ln\frac{p_\dn}{p_\up}-\gamma\hbar\Big)\Big[\n_\inl-\frac{1}{2}(\N+1)\Big].
     \label{eqn:sigma_work entire procedure}
\end{align}
An important point about this result is that the second term on the right side represents the fact that, in general, the memory is not in equilibrium with the reservoir initially---indeed, this term vanishes for $\ln\frac{p_\dn}{p_\up}=\gamma\hbar$ which corresponds to the memory and reservoir being in equilibrium initially.
Multiplying \eq{eqn:sigma_work entire procedure} by $P(\mathbf{z}^{(T)}_\fnl,\mathbf{z}^{(T)}_\inl)$ and summing over $\mathbf{z}^{(T)}_\inl$ and $\mathbf{z}^{(T)}_\fnl$ gives the total entropy production, $\ip{\sigma^{(T)}}$, which according to \eq{eqn:total entropy production > 0}, is non-negative; rearranging terms then yields
\begin{align}
    \ip{\L}
     \ge \Delta \mathcal{F}_{\rm s}^\M +\frac{1}{\gamma}\Big(\ln\frac{p_\dn}{p_\up}-\gamma\hbar\Big)\Big[p_\up-\frac{1}{2}(\N+1)\Big].
     \notag
\end{align}
Substituting the result $\Delta \mathcal{F}_{\rm s}^\M=-\frac{1}{\gamma}[\ln p_\dn -\frac{1}{2}(\N+1)(\ln \frac{p_\dn}{p_\up}-\gamma\hbar)]$, which follows from \eq{eqn:change in F_s} with \eqs{eqn:Z_i for period (1)} and \eqr{eqn:Z_f for period (2)}, gives
\begin{align}
    \ip{\L}
     &\ge -\frac{1}{\gamma}\ln p_\dn +\frac{1}{\gamma}\Big(\ln\frac{p_\dn}{p_\up}-\gamma\hbar\Big)p_\up,
\end{align}
and so for $p_\up=p_\dn=0.5$ we find
\begin{align}
    \ip{\L}
     &\ge \frac{\ln 2}{\gamma}-\frac{1}{2}\hbar.
     \label{eqn:spinlabor entire procedure}
\end{align}
This result is valid for all protocol variations, and can be compared to the variation-specific results in \eqs{eqn:Spinlabor_ineq} and \eqr{eqn:Bound 2b}.  We return to this comparison in \S\ref{sec:bounds}.

Next, we turn our attention to the spintherm cost. As no spinlabor is performed directly on the reservoir, the only way the spin angular momentum of the reservoir can change according to the first law, \eq{eqn:first_law}, is by the exchange of spintherm $\Q$ with the memory-ancilla system.  We therefore define the \emph{stochastic spintherm} absorbed by the reservoir, in analogy with the definition of stochastic heat \cite{Funo2018}, as the change in $J^\R_{z}$ along a trajectory in phase space, i.e. as
\bea   \label{eqn:stochastic spintherm}
     \Q(\mathbf{z}^{(R)}_\fnl,\mathbf{z}^{(R)}_\inl )
    &\equiv  J^\R_{z}(\mathbf{z}^\R_\fnl) -J^\R_{z}(\mathbf{z}^\R_\inl) .
\eea
Expressing only the reservoir term $\sigma^{(R)}(\mathbf{z}^{(R)}_\fnl,\mathbf{z}^{(R)}_\inl )$  in \eq{eqn:sigma} in terms of the probability distributions $f^\R_\lambda$, and then substituting for $f^\R_\lambda$ using the first factor in \eq{eqn:distribution_function} yields
\begin{align} 
    \sigma^\T(\mathbf{z}^{(T)}_\fnl,\mathbf{z}^{(T)}_\inl)
           &=\gamma \Q(\mathbf{z}^{(R)}_\fnl,\mathbf{z}^{(R)}_\inl ) +\sigma^{(M)}(\mathbf{z}^{(M)}_\fnl,\mathbf{z}^{(M)}_\inl ). \notag
\end{align}
Comparing with \eq{eqn:sigma} shows that the total stochastic entropy production is the sum of the entropy production of the memory and the entropy content $\gamma \Q(\mathbf{z}^{(R)}_\fnl,\mathbf{z}^{(R)}_\inl ) $ of the spintherm that flows into the reservoir.
As before, multiplying  by $P(\mathbf{z}^{(T)}_\fnl,\mathbf{z}^{(T)}_\inl)$ and summing over $\mathbf{z}^{(T)}_\inl$ and $\mathbf{z}^{(T)}_\fnl$ gives the total entropy production $\ip{\sigma^\T}$, and using our earlier result in \eq{eqn:total entropy production > 0}, it follows that
\bea   \label{eqn:Q > sigma}
          \gamma \ip{\Q}\ge -\ip{\sigma^{(M)}}.
\eea
We note that $\sigma^{(M)}$ is given by the last three terms of \eq{eqn:sigma_work_pre}, i.e.
\begin{align}    \label{eqn:stochastic sigma for memory}
     \sigma^{(M)}(\mathbf{z}^{(M)}_\fnl,\mathbf{z}^{(M)}_\inl )
       &=   \gamma^{(M)}_\fnl J^\M_{z}(\mathbf{z}^\M_\fnl)-\gamma^{(M)}_\inl J^\M_{z}(\mathbf{z}^\M_\inl) \notag\\
           &\hspace{-5mm}\quad +\ln \frac{Z^\M_\fnl}{Z^\M_\inl} .
\end{align}

As previously noted, initially $\mathbf{z}^\M_\inl=(\n_\inl,a_\inl)$ with $\n_\inl=0$ or $1$ with probabilities $p_\dn$ and $p_\up$, respectively, $a_\inl=0$, $\gamma^\M_{\inl}=
   \frac{1}{\hbar}\ln \Big( p_\dn / p_\up \Big)$ from \eq{eqn:gamma_M}, $Z^M_\inl$ is given by \eq{eqn:Z_i for period (1)}, and $J^\M_{z}(\mathbf{z}^\M_\inl)$ is given by \eq{eqn:JzM(z_i)}.
For the case where the maximum number of CNOT steps are performed, the values of $\n_\fnl$ in $\mathbf{z}^\M_\fnl=(\n_\fnl,a_\fnl)$ are $\n_\fnl=0$ and $1$ with probabilities $Q_\dn(\N)=1-Q_\up(\N)$ and $Q_\up(\N)$, respectively, where $Q_\up(m)$ is given in \eq{eqn:Q}, $a_\fnl=\N$, $\gamma^\M_f=\gamma$ from \eq{eqn:gamma_M}, $Z^M_\fnl$ is given by \eq{eqn:Z_f for period (2)}, and $J^\M_{z}(\mathbf{z}^\M_\fnl)$ is given by \eq{eqn:JzM(z_f)}.
Putting this all together with \eq{eqn:stochastic sigma for memory} gives
\begin{align}    \label{eqn:sigma for memory}
        \ip{\sigma^{(M)}} = \gamma Q_\up(\N) \hbar + \ln p_\downarrow - \frac{(\N + 1)}{2} \ln \frac{p_\downarrow}{p_\uparrow}
\end{align}
where we have ignored exponentially-insignificant terms of order $e^{-\frac{1}{2}(\N+1)\gamma\hbar}$.
Finally, substituting this result into \eq{eqn:Q > sigma} and setting $p_\up=p_\dn=0.5$ then shows that
\bea
    \ip{\Q} \geq \frac{\ln2}{\gamma} \label{eqn:heat bound}
\eea
as expected. This result is independent of protocol choice $C$ and can be compared with our earlier variation-dependent result in \eq{eqn:Heat_ineq}.  We return to this comparison in \S\ref{sec:bounds}.

\section{Bounds on the cost of erasure}\label{sec:bounds}

The values of $\ip{\L}_C$ and $\ip{\Q}_C$ given in \eqs{eqn:avg} and \eqr{eqn:total spintherm cost} are the average spinlabor and spintherm costs for information erasure associated with the variations of the VB protocol described in \S\ref{sec:3 protocols} under ideal conditions.  In any practical implementation, we expect losses, inefficiencies and other physical limitations to lead to higher erasure costs \cite{Croucher2020a}, and so \eqs{eqn:avg} and \eqr{eqn:total spintherm cost} represent lower bounds for the costs in this sense. This naturally raises the question of the relation between \eqs{eqn:avg} and \eqr{eqn:total spintherm cost} and the \emph{universal lower bounds} for any erasure mechanism based on expending spinlabor as spintherm.  We would also like to assess the relative merits of closed form versions of \eqs{eqn:avg} and \eqr{eqn:total spintherm cost} that we derived in previous sections.
We address these issues in this section.  We focus on the maximal-stored information case of $p_\up=p_\dn=0.5$ for brevity, leaving the extension to the general case as a straightforward exercise.

We derived the closed-form lower bound on the spinlabor cost $\ip{\L}_C$,
\bea
   \ip{\L}_C & \geq & \frac{\hbar C}{2} +\gamma^{-1} \ln(1+e^{-(C+1)\gamma \hbar}),
    \label{eqn:Bound_1}
\eea
given by \eq{eqn:Spinlabor_ineq} with $p_\up=0.5$, using an integral approximation of the sum in \eq{eqn:avg}.

\begin{figure}[H]
	\centering
	\includegraphics[width=0.48\textwidth]{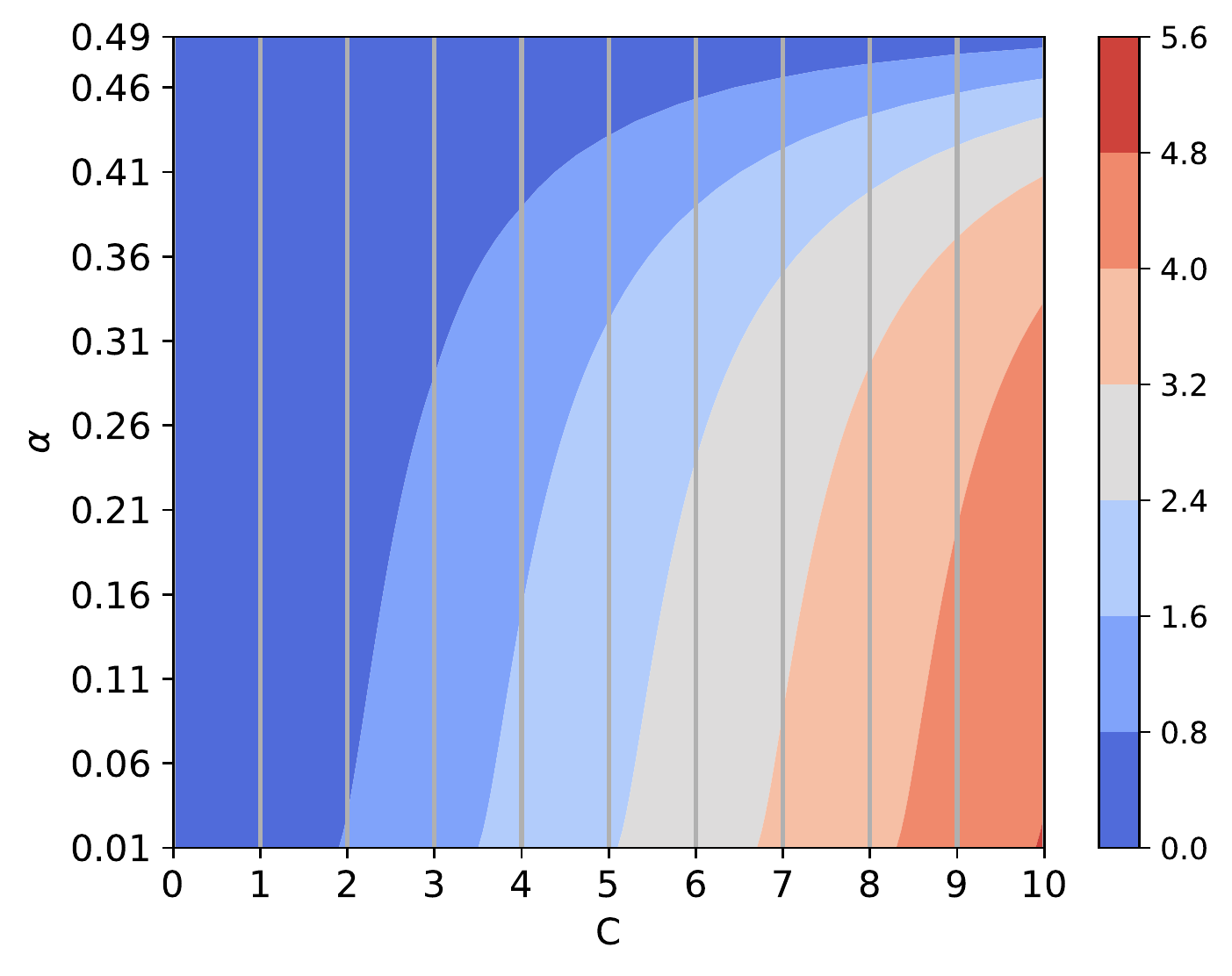}
	\caption{Plot of $\Delta B$ which compares Eq. \eqref{eqn:Bound_1} to Eq. \eqref{eqn:Bound_2} with $p_{\uparrow}=0.5$, $\alpha=0.01$ to $0.49$ and $C=0$ to $10$ discretely.}
	\label{fig:phasespace_LC_compare}
\end{figure}

We also derived a different closed-form lower bound by applying Jensen's inequality to our Jarzinsky-like equality in Eq.\eqref{eqn:Jarzynski_2} to obtain
\bea
    \langle\mathcal{L}_{\rm s}\rangle_{C} \geq  \gamma^{-1} \ln \left\lbrace \frac{2[1+e^{-(C+1)\gamma\hbar}]}{1+e^{-C\gamma\hbar}} \right\rbrace  \label{eqn:Bound_2}
\eea
as given by \eqs{eqn:Bound 2b} and \eqr{eqn:Jarzynski_2 A}. To determine which of \eqs{eqn:Bound_1} or \eqr{eqn:Bound_2} gives the tighter bound, we plot the difference $\Delta B$ between their right sides in Fig. \ref{fig:phasespace_LC_compare} as a function of reservoir spin polarization $\alpha$ and protocol variation parameter $C$, where
\begin{align}    \label{eq:Delta B}
   \Delta B &\equiv \text{RS\eqr{eqn:Bound_1}} - \text{RS\eqr{eqn:Bound_2}} \notag\\
     &=\frac{\hbar C}{2}-\gamma^{-1}\ln\left( \frac{2}{1+e^{-C\gamma\hbar}}\right)
\end{align}
and RS($X$) refers to the right side of Eq.~($X$). The lowest spinlabor cost occurs when $C=0$, for which $\Delta B=0$ indicating that both bounds on the average spinlabor cost agree. In contrast, we find that $\Delta B\to \infty$ as $C \to \infty$.  As the figure shows $\Delta B$ has only non-negative values, it clearly demonstrates that \eq{eqn:Bound_1} gives the tighter closed-form-bound overall.

This finding, however, is specific to the variations of the VB erasure protocol we have examined.  To go beyond specific erasure protocols we turn to the bound in \eq{eqn:spinlabor entire procedure} that we derived using the integral fluctuation theorem, i.e.
\begin{align}
    \ip{\L}
     &\ge \frac{\ln 2}{\gamma}-\frac{\hbar}{2}.
     \label{eqn:universal spinlabor bound}
\end{align}
Its application is limited only by deterministic evolution between the initial and final states of the memory-ancilla-reservoir system, and so it applies to every possible erasure protocol satisfying this condition. We therefore, call it the \emph{universal bound} for spinlabor expended as spintherm at inverse spin temperature $\gamma$ per bit erased.

Finally,  we show that the universal bound can be derived by lower-bounding the sum in \eq{eqn:avg} in a different way to what we did to derive \eq{eqn:Spinlabor_ineq}. Using \eq{eqn:avg}, the lowest value of spinlabor occurs for the protocol when $C=0$ and so
\bea
    \langle \mathcal{L}_{\rm s} \rangle_{C=0}
     &=& \sum_{m=0}^{\infty}\frac{\hbar e^{-m\gamma\hbar}}{1+e^{-m\gamma\hbar}} - \frac{\hbar}{2}. \label{eqn:total mean cost czero}
\eea
where we have adjusted the summation index $m$ and lower limit to include an extra term equal to $\frac{1}{2}\hbar$. The sum on the right side is bounded as follows
\begin{align*}
    \sum_{m=0}^{\infty} \frac{\hbar e^{-m\gamma\hbar}}{1+e^{-m\gamma\hbar}} \geq
    \int_{m=0}^{\infty} \frac{\hbar e^{-m\gamma\hbar}}{1+e^{-m\gamma\hbar}}dm  \geq \frac{\ln 2}{\gamma} ,
\end{align*}
and so we find that the average spinlabor cost is bounded by
\bea   \label{eqn:approx uni spinlabor bd}
    \ip{\mathcal{L}_{\rm s}}_{C=0}
    \geq \frac{\ln 2}{\gamma} - \frac{1}{2}\hbar
\eea
in agreement with the universal bound in \eq{eqn:universal spinlabor bound}.  
We have already noted that the spinlabor cost is lowest for the protocol with $C=0$, i.e. $\ip{\mathcal{L}_{\rm s}}_{C}>\ip{\mathcal{L}_{\rm s}}_{0}$ for $C>0$, which suggests that  larger values of $C$ give tighter bounds on the spinlabor cost.
Indeed, it is straightforward to show graphically that
\begin{align}
      \frac{\hbar C}{2} +\gamma^{-1} \ln(1+e^{-(C+1)\gamma \hbar}) >   \frac{\ln 2}{\gamma} - \frac{1}{2}\hbar
\end{align}
for all values of $\gamma>0$ and $C\ge 0$, and so \eq{eqn:Bound_1} gives a tighter bound on the spinlabor cost for the protocol variation with $C\ge 0$ compared to the universal bound \eq{eqn:universal spinlabor bound}.

The situation for the spintherm cost follows immediately from \eq{eqn:Heat_ineq} with $p_\up=0.5$, i.e. 
\begin{align}    \label{eqn:spintherm bound}
     \ip{\Q}_C &\geq \frac{(C+1) \hbar}{2} +\gamma^{-1} \ln(1+e^{-(C+1)\gamma \hbar}).
\end{align}
which is the tightest closed-form bound we have for variations of the VB erasure protocol.
Moreover, the spintherm bound in \eq{eqn:heat bound} that we derived using the integral fluctuation theorem, i.e.
\begin{align}    \label{eqn:universal spintherm bound}
    \ip{\Q} &\geq \frac{\ln2}{\gamma}  , 
\end{align}
like \eq{eqn:universal spinlabor bound}, applies to every possible erasure protocol with deterministic evolution, and so we call it the \emph{universal bound} for spintherm transferred to the reservoir at inverse spin temperature $\gamma$ per bit erased. Nevertheless, according to the foregoing discussion of the spinlabor cost, \eq{eqn:spintherm bound} gives a tighter bound on the spintherm cost for protocol variation $C$ compared to \eq{eqn:universal spintherm bound}.

\section{Conclusion} \label{sec:conclusion}

In conclusion, we have extended our earlier study \cite{Croucher2017} of the discrete fluctuations and average bounds of the erasure cost in spin angular momentum for Vaccaro and Barnett's proposed information erasure protocol \cite{Vaccaro2011,Barnett2013}. We generalized the protocol to include multiple variations characterized by the number $C$ of CNOT operations that have been performed on the memory-ancilla system before it is first brought into equilibrium with the spin reservoir. We also clarified the erasure costs in terms of the spin equivalent of work, called spinlabor, and the spin equivalent of heat, called spintherm.  We showed that the previously-found bound on the erasure cost of $\gamma^{-1}\ln{2}$ can be violated by the spinlabor cost, and only applies to the spintherm cost. We derived a Jarzynski equality and an integral fluctuation theorem associated with spin reservoirs, and applied them to analyze the costs of information erasure for the generalized protocols. Finally we derived a number of bounds on the spinlabor and spintherm costs, including closed-form approximations, and determined the tightest ones. 

This work is important for the design and implementation of new kinds of heat engines and batteries that use multiple conserved quantities, particularly if the quantities are discrete. The analysis of the probability of violation is crucial in the understanding of the statistics and the relation to the fluctuation theorem. In addition, it also clarifies the need for different bounds for the spinlabor and spintherm costs. This difference occurs due to the discrete nature of the conserved quantity. Work in preparation investigates the consequence of a finite spin reservoir \cite{Croucher2020a}. Other future work within this field may look into quantum energy teleportation (QET) and how this improved algorithmic cooling method can be applied to extract entropy from the qubit (memory) more efficiently \cite{Briones2017}.

\section*{Acknowledgements}
This research was supported by the ARC Linkage Grant No. LP180100096 and the Lockheed Martin Corporation. TC acknowledges discussions with S. Bedkihal. We acknowledge the traditional owners of the land on which this work was undertaken at Griffith University, the Yuggera people.

\vspace{3mm}
\appendix

\section{\label{sec:ap anal prob soln} Analytical expression for \texorpdfstring{$\P_{m}(n)$}{Pm(n)}}
\vspace{-3mm}

In this Appendix we derive an analytical expression for $\P_{m}(n)$, the probability for the accumulated spinlabor cost of $n\hbar$ after $m$ ancilla CNOT operations, as defined by \eqs{eqn:secondeq m=0}-\eqr{eqn:recrel}.
We use the recurrence relation Eq. \eqref{eqn:recrel} to express  $\P_{C+j}(n)$ for $j>0$ in terms of the initial values $\{\P_{C}(k):0\le k\le C\}$, where $C$ is the number of ancilla CNOT operations performed before the first equilibration step.
There are two different sets of initial values, depending on the value of $C$.
According to \eq{eqn:secondeq m=0}, if $C=0$ the initial values are
\bea   \label{eqn:ap initial values P for C=0}
     \P_{0}(n) =\left\{\begin{array}{l}  1 \text{ for }n=0\\ 0 \text{ for } n>0   \end{array}   \right.
\eea
whereas according to \eq{eqn:secondeq}, if $C>0$ they are
\bea   \label{eqn:ap initial values P for C>0 and 0< n < C}
     \P_{C}(n) =\left\{\begin{array}{l}  p_\dn \text{ for }n=0\\
                                0 \text{ for } 0<n<C\\
                                p_\up \text{ for }n=C. \end{array}   \right.
\eea
For convenience, we set $\P_{m}(n)=0$ for $n<0$, and define
\begin{align}  \label{eqn:ap definitions R and S}
       R \equiv e^{-\gamma \hbar}, \qquad S_m\equiv \frac{1}{1+e^{-m\gamma \hbar}}
\end{align}
to produce a more compact notation in which \eq{eqn:Q} becomes
\begin{align*}
      Q_{\dn}(m)&=S_{m+1}, \qquad
      Q_{\up}(m)=R^{m+1}S_{m+1}
\end{align*}
and the recurrence relation \eq{eqn:recrel} reduces to
\begin{align}
    \P_{m}(n)&=\left[\P_{m-1}(n) +\P_{m-1}(n-1)R^{m}\right]S_m .
    \label{eqn:ap recurrence reln}
\end{align}
We immediately find from applying \eq{eqn:ap recurrence reln} recursively that
\begin{widetext}
\begin{align}
    \P_{C+j}(n)&=[\P_{C+j-1}(n)+\P_{C+j-1}(n-1)R^{C+j}]S_{C+j} , \notag\\
         &=\left[\P_{C+j-2}(n)+\P_{C+j-2}(n-1)\sum_{\ell=0}^{1}R^{C+j-\ell}
         +\P_{C+j-2}(n-2)\sum_{k=1}^{1}\sum_{\ell=0}^{k-1}R^{C+j-k}R^{C+j-\ell}\right] \prod_{\ell=0}^{1} S_{C+j-\ell}\notag\\
         &=\left[\P_{C+j-3}(n)+\P_{C+j-3}(n-1)\sum_{\ell=0}^{2}R^{C+j-\ell}
         +\P_{C+j-3}(n-2)\sum_{k=1}^{2}\sum_{\ell=0}^{k-1}R^{C+j-k}R^{C+j-\ell}\right.\notag\\
         &\qquad \left.+\P_{C+j-3}(n-3)\sum_{i=2}^{2}\sum_{k=1}^{i-1}\sum_{\ell=0}^{k-1}R^{C+j-i}R^{C+j-k}R^{C+j-\ell}\right] \prod_{\ell=0}^{2} S_{C+j-\ell}.\notag
\end{align}
We are interested in the large-$j$ limit, and so we need only consider $j>n$ for any given value of $n$, in which case the recursion leads eventually to
\begin{align} \label{eqn:ap general expression for P}
    \P_{C+j}(n)&=\left[\P_{C}(n)+\P_{C}(n-1)\sum_{\ell=0}^{j-1}R^{C+j-\ell}
         +\P_{C}(n-2)\sum_{k=1}^{j-1}\sum_{\ell=0}^{k-1}R^{C+j-k}R^{C+j-\ell}\right.\notag\\
         &\qquad \left.+\dotsb+\P_{C}(0)\underbrace{\sum_{m=n-1}^{j-1}\dotsb\sum_{i=2}^{\dotsb}\sum_{k=1}^{i-1}\sum_{\ell=0}^{k-1}} \;\underbrace{\text{\raisebox{0mm}[0mm][4.75mm]{}}R^{C+j-m}\dotsb R^{C+j-i}R^{C+j-k}R^{C+j-\ell}}\right] \prod_{\ell=0}^{j-1} S_{C+j-\ell}.\\
         &\hspace{3.75cm}\text{\small $n$ nested sums}\hspace{2.5cm}\text{\small $n$ factors}\notag
\end{align}
We call the set of multiple sums ``nested'' because, except for the leftmost sum, the limits of each sum is related to the neighboring sum on its left in that the lower limit ($\ell=0$ for the last sum) is one less than the neighboring lower limit ($k=1$) and the upper limit ($\ell=k-1$) is one less the value of the neighboring summation index ($k$, respectively).
This general result simplifies considerably when evaluated for cases with specific ranges of values.

Case (\emph{i}) corresponds to $C=0$ and $j>n$, and so the probabilities on the right side of \eq{eqn:ap general expression for P} are given by \eq{eqn:ap initial values P for C=0}.
Thus, only the last term in square brackets in \eq{eqn:ap general expression for P} survives, and so
\begin{align}   \label{eqn:ap P_j(n) for C=0}
    \P_{j}(n)&=A(j,n)R^{nj}\prod_{\ell=0}^{j-1} S_{j-\ell}
\end{align}
where we have defined
\begin{align}    \label{eqn:ap A(r,q) definition}
        A(j,n)&\equiv\underbrace{\sum_{m=n-1}^{j-1}\dotsb\sum_{i=2}^{\dotsb}\sum_{k=1}^{i-1}\sum_{\ell=0}^{k-1}} \;\underbrace{\text{\raisebox{0mm}[0mm][4.75mm]{}}R^{-m}\dotsb R^{-i}R^{-k}R^{-\ell}}
        =\prod_{k=0}^{n-1}\frac{R^{-k}-R^{-j}}{1-R^{-(k+1)}}\\
         &\hspace{1.3cm}\text{\small $n$ nested sums}\hspace{1.3cm}\text{\small $n$ factors}\notag
\end{align}
for integers $j\ge n>0$ and set $A(j,0)\equiv 1$, and we have used \eq{eqn:ap reduced nested sums} from \hyperref[sec:ap reducing nested sums]{Appendix \ref{sec:ap reducing nested sums}} to derive the expression on the far right of \eq{eqn:ap A(r,q) definition}.

Case (\emph{ii}) corresponds to $C>0$ and $j>n$. In this case we use \eq{eqn:ap initial values P for C>0 and 0< n < C} to replace $\P_C(k)$ for $k=0,1,2,\dotsc,n$ on the right side of \eq{eqn:ap general expression for P} to find
\begin{align}   \label{eqn:ap P_j(n) for C>0 and n<C}
    \P_{C+j}(n)&=p_\dn A(j,n)R^{n(C+j)}\prod_{\ell=0}^{j-1} S_{j-\ell}
\end{align}
for $n<C$, and
\begin{align}   \label{eqn:ap P_j(n) for C>0 and n>C}
    \P_{C+j}(n)&=\Big(p_\dn A(j,n)R^{n(C+j)} + p_\up A(j,n-C)R^{(n-C)(C+j)}\Big)\prod_{\ell=0}^{j-1} S_{j-\ell}
\end{align}
for $n\ge C$.
Interestingly, substituting $C=0$ into \eq{eqn:ap P_j(n) for C>0 and n>C} and using $p_\up+p_\dn=1$ gives the same result as \eq{eqn:ap P_j(n) for C=0} for case (\emph{i}).

As the cycles of the ancilla CNOT step followed by the equilibration step are repeated indefinitely, the statistics of a complete erasure process corresponds to the limit $j\to\infty$.
Substitution and rearranging using \eqs{eqn:ap definitions R and S} and \eqr{eqn:ap A(r,q) definition} gives the following limiting values,
\begin{align}    \label{eqn:ap limiting values}
     \lim_{j\to\infty}\prod_{\ell=0}^{j-1} S_{j-\ell}
         &=\lim_{j\to\infty}\prod_{\ell=0}^{j-1} S_{\ell+1}
           =\lim_{j\to\infty}\prod_{\ell=0}^{j-1} \frac{1}{1+e^{-(\ell+1)\gamma \hbar}}
           =\frac{1}{(-e^{-\gamma \hbar};e^{-\gamma \hbar})_\infty},\\
     \lim_{j\to\infty}A(j,n)R^{nj}
         &=\lim_{j\to\infty}\prod_{k=0}^{n-1}\Big(\frac{R^{-k}-R^{-j}}{1-R^{-(k+1)}}R^{j}\Big)
           =\lim_{j\to\infty}\prod_{k=0}^{n-1}\frac{e^{(k-j)\gamma\hbar}-1}{1-e^{(k+1)\gamma\hbar}}
           =\prod_{k=0}^{n-1}\frac{ e^{-(k+1)\gamma\hbar}}{1-e^{-(k+1)\gamma\hbar}} \notag\\
          &=\frac{ e^{-\frac{1}{2}n(n+1)\gamma\hbar}}{(e^{-\gamma\hbar};e^{-\gamma\hbar})_n},\\
     \lim_{j\to\infty}A(j,n)R^{n(C+j)}
         &=\lim_{j\to\infty}R^{nC}\prod_{k=0}^{n-1}\Big(\frac{R^{-k}-R^{-j}}{1-R^{-(k+1)}}R^{j}\Big)
           =\frac{ e^{-n(C+\frac{n+1}{2})\gamma\hbar}}{(e^{-\gamma\hbar};e^{-\gamma\hbar})_n},\\
     \lim_{j\to\infty}A(j,n-C)R^{(n-C)(C+j)}
         &=\lim_{j\to\infty}R^{(n-C)C}\prod_{k=0}^{n-C-1}\Big(\frac{R^{-k}-R^{-j}}{1-R^{-(k+1)}}R^{j}\Big)
           =\frac{ e^{-(n-C)(C+\frac{n-C+1}{2})\gamma\hbar}}{(e^{-\gamma\hbar};e^{-\gamma\hbar})_{n-C}},
\end{align}
where $(a;q)_n$ is the $q$-Pochhammer symbol
\begin{align}    \label{eqn:ap pochhammer}
        (a;q)_n\equiv\prod_{k=0}^{n-1}(1-a q^k), \quad (a;q)_0\equiv 1.
\end{align}
Using these results together with \eqs{eqn:ap P_j(n) for C=0}, \eqr{eqn:ap P_j(n) for C>0 and n<C} and \eqr{eqn:ap P_j(n) for C>0 and n>C} gives the probability for a spinlabor cost of $n\hbar$ for the full erasure procedure in case (\emph{i}), i.e. $C=0$, as
\begin{align}    \label{eqn:ap P_infty for C=0}
        \P_\infty(n)=\frac{ e^{-\frac{1}{2}n(n+1)\gamma\hbar}}{(e^{-\gamma\hbar};e^{-\gamma\hbar})_n(-e^{-\gamma \hbar};e^{-\gamma \hbar})_\infty}
\end{align}
and in case (\emph{ii}), i.e. $C>0$, as
\begin{align}    \label{eqn:ap P_infty for C>0}
    \P_{\infty}(n)&=\left\{\ary{ll}{
        p_\dn\frac{e^{-n(C+\frac{n+1}{2})\gamma\hbar}}{(e^{-\gamma\hbar};e^{-\gamma\hbar})_n(-e^{-\gamma \hbar};e^{-\gamma \hbar})_\infty} , \text{ for }n<C \\
        p_\dn\frac{e^{-n(C+\frac{n+1}{2})\gamma\hbar}}{(e^{-\gamma\hbar};e^{-\gamma\hbar})_n(-e^{-\gamma \hbar};e^{-\gamma \hbar})_\infty}
           + p_\up \frac{ e^{-(n-C)(C+\frac{n-C+1}{2})\gamma\hbar}}{(e^{-\gamma\hbar};e^{-\gamma\hbar})_{n-C}(-e^{-\gamma \hbar};e^{-\gamma \hbar})_\infty}, \text{ for }n\ge C.    }
                  \right.
\end{align}

\section{\label{sec:ap reducing nested sums} Reducing the nested sums}

Here we reduce the expression for $A(j,n)$ in \eq{eqn:ap A(r,q) definition} using a technique introduced by one of us in a different context \cite{Vaccaro2011a}.
It is convenient to consider the $n$-fold nested sums of the form
\begin{align}
      \sum_{k=n-1}^{j-1}\sum_{\ell=n-2}^{k-1}\sum_{m=n-3}^{\ell-1}\!\!\cdots\sum_{p=1}^{\dotsc}
      \sum_{q=0}^{p-1} r^{k+\ell+m+\cdots+p+q}
\end{align}
for $r=R^{-1}$ and given values of $j$ and $n$.
Changing the order in which the indices $k$ and $\ell$ are summed, we find
\begin{align}
      \sum_{k=n-1}^{j-1}\sum_{\ell=n-2}^{k-1}\sum_{m=n-3}^{\ell-1}\!\!\cdots\sum_{p=1}^{\dotsc}
      \sum_{q=0}^{p-1} r^{k+\ell+m+\cdots+p+q}= \sum_{\ell=n-2}^{j-2}\sum_{k=\ell+1}^{j-1}\sum_{m=n-3}^{\ell-1}\!\!\cdots\sum_{p=1}^{\dotsc}\sum_{q=0}^{p-1}
      r^{k+\ell+m+\cdots+p+q}\ ,
\end{align}
next, by cyclically interchanging the indices in the order $k\to q\to p\to o\to\cdots\to
m\to\ell\to k$ on the right-hand side, we get
\begin{align}
     \sum_{k=n-1}^{j-1}\sum_{\ell=n-2}^{k-1}\sum_{m=n-3}^{\ell-1}\!\!\cdots\sum_{p=1}^{\dotsc}
      \sum_{q=0}^{p-1} r^{k+\ell+m+\cdots+p+q}
      = \sum_{k=n-2}^{j-2}\sum_{q=k+1}^{j-1}\sum_{\ell=n-3}^{k-1}\!\!\cdots\sum_{o=1}^{\dotsc}\sum_{p=0}^{o-1}
      r^{q+k+\ell+\cdots+o+p}\ ,
\end{align}
and finally, bringing the sum over $q$ to the extreme right on the right-hand side and rearranging gives
\begin{align}
     \sum_{k=n-1}^{j-1}\sum_{\ell=n-2}^{k-1}\sum_{m=n-3}^{\ell-1}\!\!\cdots\sum_{p=1}^{\dotsc}
      \sum_{q=0}^{p-1} r^{k+\ell+m+\cdots+p+q}
      = \sum_{k=n-2}^{j-2}\sum_{\ell=n-3}^{k-1}\sum_{m=n-4}^{\ell-1}\!\!\cdots\sum_{p=0}^{\dotsc}\sum_{q=k+1}^{j-1}
      r^{k+\ell+m+\cdots+p+q}\ .
\end{align}
We abbreviate this general summation property as
\begin{align}
     \sum_{k=n-1}^{j-1}\cdots\sum_{o=2}^{\dotsc}\sum_{p=1}^{o-1}\sum_{q=0}^{p-1} r^{k+\cdots+o+p+q}
      = \sum_{k=n-2}^{j-2}\cdots\sum_{s=1}^{\dotsc}\sum_{p=0}^{o-1}\sum_{q=k+1}^{j-1}
      r^{k+\cdots+o+p+q}\ .
      \label{general sum prop}
\end{align}
Consider the product
\begin{align}
     \Big( r+1\Big)\sum_{p=1}^{o-1}\sum_{q=0}^{p-1} r^{p+q}
      &=r\left[\sum_{p=0}^{o-2}\sum_{q=p-1}^{o-1}\!\! r^{p+q}\right]+\sum_{p=1}^{o-1}\sum_{q=0}^{p-1}
      r^{p+q}
      =\sum_{p=1}^{o-1}\sum_{q=p}^{o-1} r^{p+q}+\sum_{p=1}^{o-1}\sum_{q=0}^{p-1}
      r^{p+q}
      =\sum_{p=1}^{o-1}\sum_{q=0}^{o-1} r^{p+q}
      \label{2 sums}
\end{align}
where we have used \eq{general sum prop} to rearrange the sums in the square bracket.
The two {\it nested} summations on the far left have been reduced to two {\it
un-nested} summations on the far right. Similarly,
\begin{align}
     \Big( r^{2}\!\!+\!r\!\!+\!1\Big)\sum_{o=2}^{n-1}\sum_{p=1}^{o-1}\sum_{q=0}^{p-1} r^{o+p+q}
     &= r^{2}\left[\sum_{o=1}^{n-2}\sum_{p=0}^{o-1}\sum_{q=o+1}^{n-1} r^{o+p+q}\right]
                  +\sum_{o=2}^{n-1}\Big(r+1\Big)\sum_{p=1}^{o-1}\sum_{q=0}^{p-1}
     r^{o+p+q}\notag\\
     &= \sum_{o=2}^{n-1}\sum_{p=1}^{o-1}\sum_{q=o}^{n-1} r^{o+p+q}
                  +\sum_{o=2}^{n-1}\left[\sum_{p=1}^{o-1}\sum_{q=0}^{o-1} r^{o+p+q}\right]
                  = \sum_{o=2}^{n-1}\sum_{p=1}^{o-1}\sum_{q=0}^{n-1} r^{o+p+q}
\end{align}
where \eq{general sum prop} and \eq{2 sums} have been used to derive the terms in square brackets,
{\it three} nested summations on the far left side have been reduced to {\it
two} nested summations and one un-nested summation on the far right side.
It follows that for $n$ nested sums,
\begin{align}
     &\Big(\sum_{\ell=0}^{n-1}r^{\ell}\Big)\!\!\underbrace{\ \sum_{m=n-1}^{j-1}\!\!\cdots\sum_{o=2}^{\dotsc}\sum_{p=1}^{o-1}\sum_{q=0}^{p-1}\
     }\!\!r^{m+\cdots+o+p+q}
= \Big(\!\!\underbrace{\ \sum_{m=n-1}^{j-1}\!\!\cdots\sum_{o=2}^{\dotsc}\sum_{p=1}^{o-1}\ }\!\!
      r^{m+\cdots+o+p}\Big)\sum_{q=0}^{j-1} r^{q}\ .\label{eqn:ap p nested sums}\\
      &\hspace{1.95cm}\mbox{\small $n$ nested sums}\hspace{3.25cm}\mbox{\small $n\!-\!1$ nested sums}\notag
\end{align}
Consider repeating this calculation for the $n-1$ nested sums on the right side, i.e.
\begin{align}
     &\Big(\sum_{\ell=0}^{n-2}r^{\ell}\Big)\!\!\underbrace{\ \sum_{m=n-1}^{j-1}\!\!\cdots\sum_{o=2}^{\dotsc}\sum_{p=1}^{o-1}\ }
      r^{m+\cdots+o+p}
      =\Big(\sum_{\ell=0}^{n-2}r^{\ell}\Big)r^{n-1}\!\!\sum_{m=n-2}^{j-2}\!\!\cdots\sum_{o=1}^{\dotsc}\sum_{p=0}^{o-1}
      r^{m+\cdots+o+p}
      =\Big(\!\!\underbrace{\sum_{m=n-1}^{j-1}\!\!\cdots\sum_{o=2}^{\dotsc}\ }
      r^{m+\cdots+o}\Big)\sum_{p=1}^{j-1}r^{p}\notag\\
      &\hspace{1.45cm}\mbox{\small $n\!-\!1$ nested sums}\hspace{8.65cm}\mbox{\small $n\!-\!2$ nested sums}\notag
\end{align}
where we temporarily factored out $r^{n-1}$ in the intermediate expression by redefining each summation variables to be one less in value, and used \eq{eqn:ap p nested sums} to arrive at the final result.
Thus, $n$ iterations of this calculation yields
\begin{align}
     \prod_{k=0}^{n-1}\!\Big(\sum_{\ell=0}^{k}r^{\ell}\Big) \sum_{m=n-1}^{j-1}\!\cdots\!\sum_{o=2}^{\dotsc}\sum_{p=1}^{o-1}\sum_{q=0}^{p-1}
                        r^{m+\cdots+o+p+q}
      &=\prod_{k=0}^{n-1}\Big(\sum_{\ell=k}^{j-1} r^{\ell}\Big),
\end{align}
and so
\begin{align}  \label{eqn:ap reduced nested sums}
     \sum_{m=n-1}^{j-1}\cdots\sum_{o=2}^{\dotsc}\sum_{p=1}^{o-1}\sum_{q=0}^{p-1}
                        r^{m+\cdots+o+p+q}
      &=\frac{\prod_{k=0}^{n-1}\Big(\sum_{\ell=k}^{j-1}
      r^{\ell}\Big)}{\prod_{k=0}^{n-1}\Big(\sum_{\ell=0}^{k}r^{\ell}\Big)}
      =\prod_{k=0}^{n-1}\frac{r^{k}-r^{j}}{1-r^{k+1}},
\end{align}
where we have evaluated two geometric series in arriving at the last expression.
\end{widetext}

\section{Gaussian distribution as \texorpdfstring{$\alpha\to 0.5$}{alpha to 0.5} \label{sec:ap gaussian distribution}}

Fig.~\ref{fig:work_dist} shows that the spinlabor distribution $Pr(\L)$ is Gaussian-like for $\alpha=0.4$ and raises the question whether it approaches a Gaussian distribution as $\alpha\to 0.5$. We address this question here.
Recall from \eq{eqn:gamma} that $\alpha\to 0.5$ implies $\gamma\to 0$.
A rough estimate of the nature of $Pr(\L)$ in this limit can be found by approximating both $Q_\up(m)$ and $Q_\dn(m)$ with $0.5$, which is their limiting value as $\gamma\to 0$ according to \eq{eqn:Q}.
This entails approximating the recurrence relation \eq{eqn:recrel} for $m>C$ with
\begin{align}    \label{eqn:ap recur approx}
    \P_{m}(n)&\approx \frac{1}{2}[\P_{m-1}(n)+\P_{m-1}(n-1)],
\end{align}
which yields
\begin{align*}
    &\P_{m+1}(n)\approx \frac{1}{2}[\P_{m}(n)+\P_{m}(n-1)]  \\
    &\qquad\approx \frac{1}{2^2}[\P_{m-1}(n)+2\P_{m-1}(n-1)+\P_{m-1}(n-2)],
\end{align*}
on one iteration of \eq{eqn:ap recur approx}, and
\begin{align}   \label{eqn:ap prob approx for gamma 0}
    \P_{m+k}(n)\approx \frac{1}{2^{k+1}}\sum_{j=0}^{k+1}\binom{k+1}{j}\P_{m-1}(n-j),
\end{align}
on $k$, due to its binary-tree structure, where $\binom{\boldsymbol{\cdot}}{\boldsymbol{\cdot}}$ is the binomial coefficient symbol. Treating the $C=0$ case, setting $m=1$ and adjusting the value of $k$ yields
\begin{align}
    \P_{k}(n)\approx \frac{1}{2^{k}}\sum_{j=0}^{k}\binom{k}{j}\P_{0}(n-j),
\end{align}
which becomes
\begin{align}
    \P_{k}(n)\approx \frac{1}{2^{k}}\binom{k}{n}
\end{align}
according to \eq{eqn:secondeq m=0} provided $k>n$, and thus
\begin{align}   \label{eqn:ap divergent gaussian}
    \P_{k}(n)\approx \frac{1}{\sqrt{\sfc{1}{2}k\pi}}\exp\left[-\frac{1}{\sfc{1}{2}k}(n-\sfc{1}{2}k)^2\right]
\end{align}
using the Gaussian approximation to a binomial distribution.
Although the Gaussian nature is clearly evident, the difficulty with this rough calculation is that the mean spinlabor cost of $\ip{\L}=\sum_n \P_k(n)n\hbar \approx\sfc{1}{2}k\hbar$ diverges with the number of CNOT steps $k$.

\begin{figure}[H]
\centering
\includegraphics[width=0.48\textwidth]{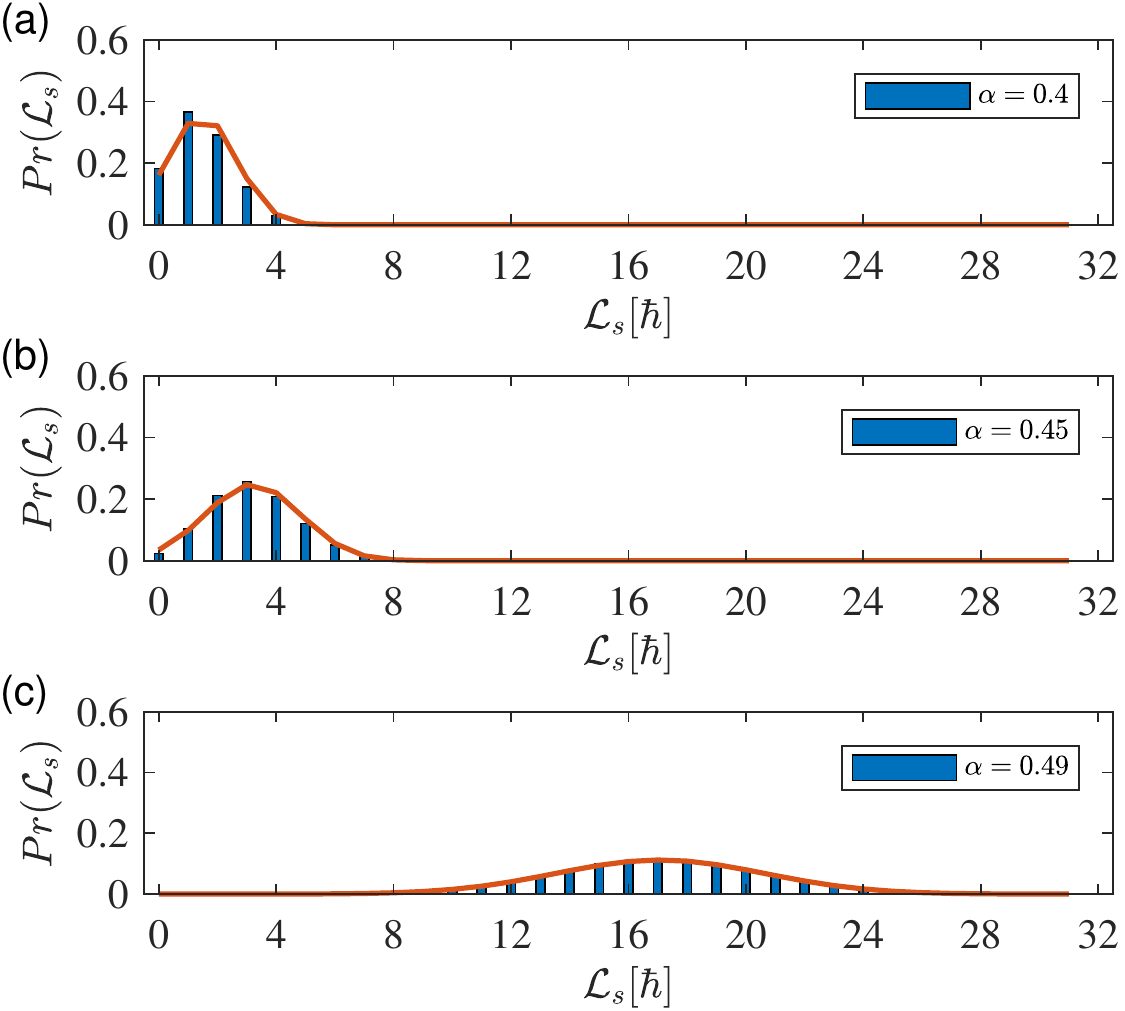}
\caption{Spinlabor distribution with an overlapping gaussian curve at $C=0$.}
  \label{fig:gaussian}
\end{figure}

A more convincing demonstration of the Gaussian nature is given by a direct graphical comparison with a Gaussian distribution of the same average and variance. It is shown in Fig \ref{fig:gaussian} that if $\alpha$ is close to $0.5$ the spinlabor distribution becomes close to a gaussian distribution.

\section{Average and Variance for spinlabor  \label{sec:ap average and variance}}

\vspace{-5mm}
The average spinlabor cost after $m$ CNOT steps is given by
\begin{align}  \label{eqn:ap spinlabor m}
    {[\ip{\L}_{C}]}_m &= \sum_{n=0}^{m} n\hbar \P_{m}(n)
\end{align}
where the bracket symbol $[\cdot]_m$ indicates the enclosed value is for the first $m$ CNOT steps, and thus is for an incomplete erasure process.  In deciding the summation limits in \eq{eqn:ap spinlabor m}, we used the fact that $\P_m(n)$ is zero for $n<0$ and $n>m$ because the CNOT does not extract spinlabor and the maximum spinlabor cost from $m$ CNOT steps is $m\hbar$.
Turning our attention to the case $m>C$ for which we can use the recurrence relation \eq{eqn:recrel}, we find
\begin{widetext}
\begin{align}
   {[\ip{\L}_{C}]}_m &= \sum_{n=0}^{m} n\hbar \Big[\P_{m-1}(n)Q_{\dn} (m-1)
                    +\P_{m-1}(n-1)Q_{\up} (m-1)\Big] \nonumber\\
          &= \sum_{n=0}^{m-1} n\hbar \P_{m-1}(n)Q_{\dn}(m-1)
                    +\sum_{n=0}^{m-1} (n+1)\hbar \P_{m-1}(n)Q_{\up}(m-1) \nonumber\\
          &= \sum_{n=0}^{m-1} n\hbar \P_{m-1}(n)\Big[Q_{\dn}(m-1)+Q_{\up}(m-1)\Big]
                    +\sum_{n=0}^{m-1} \hbar \P_{m-1}(n)Q_{\up}(m-1) \nonumber\\
          &= {[\ip{\L}_{C}]}_{m-1}+\hbar Q_{\up}(m-1) \label{eqn:ap average spinlabor m = m-1}
\end{align}
\end{widetext}
where we have suitably adjusted the summation limits, rearranged expressions, and used the facts that $Q_{\dn}(m-1)+Q_{\up}(m-1)=1$ and $\sum_{n=0}^{m} \P_{m}(n)=1$.
\eq{eqn:ap average spinlabor m = m-1} is a recurrence relation with respect to the index $m$.  Iterating over it once gives
\begin{align}
   {[\ip{\L}_C]}_{m} &= {[\ip{\L}_C]}_{m-2}+\hbar Q_{\up}(m-1)+\hbar Q_{\up}(m-2),
\end{align}
and thus iterating $m-1-C$ times, leads to
\begin{align}  \label{eqn:ap average spinlabor m = sum Q}
   {[\ip{\L}_C]}_{m} &={[\ip{\L}_C]}_{C}+\hbar \sum_{n=C}^{m-1}Q_{\up}(n).
\end{align}
No further iterations are possible because we have reached the initial value of the recurrence relation.
The value of ${[\ip{\L}_{C}]}_C$ is the spinlabor cost before the first equilibration step and is calculated at the beginning of {\S\ref{sec:spinlabor} to be $C\hbar p_\up$.  The cost of a full erasure procedure, obtained in the limit $m\to \infty$, is thus
\begin{align}  \label{eqn:ap average spinlabor}
   \ip{\L}_C &=C \hbar p_\up + \hbar \sum_{n=C}^{\infty}Q_{\up}(n).
\end{align}
This result can be expressed in a closed form as
\begin{align}  \label{eqn:ap average spinlabor - closed}
   \ip{\L}_C &=C \hbar p_\up + \Lambda(\gamma,C)
\end{align}
where we have defined
\begin{align}    \label{eqn:ap Lambda definition}
      \Lambda(\gamma,C)\equiv   -\frac{\psi_q(z)+\ln(1-q)}{\ln q}
\end{align}
with $q=e^{-\gamma\hbar}$, $z=i\pi-(C+1)\gamma\hbar$ and
\begin{align}
    \psi_q(z)\equiv -\ln(1-q)+\ln q\sum_{n=0}^{\infty}\frac{q^{n+z}}{1-q^{n+z}}
\end{align}
is the $q$-digamma function \cite{WeissteinLambert}, however, the closed form does not appear to have any advantages over the basic result \eq{eqn:ap average spinlabor}, and so we shall not use it in the following.

The variance in the spinlabor after $m$ CNOT steps,
\bea   \label{eqn:ap Var(L)_m}
    {[\Var(\L)_C]}_m &=& {[\ip{\L^2}_C]}_m-{[\ip{\L}_C]}_m^2,
\eea
is calculated in a similar manner. Using the recurrence relation \eq{eqn:recrel} and the method that led to \eq{eqn:ap average spinlabor m = m-1}, we find
\begin{widetext}
\bea
    {[\ip{\L^2}_C]}_m &=& \sum_{n=0}^m (n\hbar)^{2} \P_{m}(n)
  = \sum_{n=0}^{m} (n\hbar)^{2} \Big[\P_{m-1}(n)Q_{\dn}(m-1)
                    +\P_{m-1}(n-1)Q_{\up}(m-1)\Big] \nonumber \\
  &=& \sum_{n=0}^{m-1} (n\hbar)^{2}\P_{m-1}(n)Q_{\dn} (m-1)
                    +\sum_{n=0}^{m-1}[(n+1)\hbar]^{2}\P_{m-1}(n)Q_{\up}(m-1) \nonumber \\
  &=& \sum_{n=0}^{m-1} (n\hbar)^{2}\P_{m-1}(n)\Big[Q_{\dn}(m-1)+Q_{\up}(m-1)\Big]
                    +\sum_{n=0}^{m-1}(2n+1)\hbar^{2}\P_{m-1}(n)Q_{\up}(m-1) \nonumber \\
  &=& {[\ip{\L^2}_C]}_{m-1} + 2\hbar{[\ip{\L}_C]}_{m-1}Q_{\up}(m-1)+\hbar^2Q_{\up}(m-1),
\eea
which is a recurrence relation with respect to the index $m$. Iterating it once yields
\bea
    {[\ip{\L^2}_C]}_m &=& {[\ip{\L^2}_C]}_{m-2} + 2\hbar\sum_{n=m-2}^{m-1}{[\ip{\L}_C]}_{n}Q_{\up}(n)+\hbar^2\sum_{n=m-2}^{m-1}Q_{\up}(n),
\eea
and $m-1-C$ times yields
\bea   \label{eqn:ap L_s^2_m}
    {[\ip{\L^2}_C]}_m &=& {[\ip{\L^2}_C]}_{C} + 2\hbar\sum_{n=C}^{m-1}{[\ip{\L}_C]}_{n}Q_{\up}(n)+\hbar^2\sum_{n=C}^{m-1}Q_{\up}(n),
\eea
Combining this with \eqs{eqn:ap average spinlabor m = sum Q} and \eqr{eqn:ap Var(L)_m} gives
\bea
    {[\text{Var}(\L)_C]}_m &=& {[\ip{\L^2}_C]}_{C} + 2\hbar\sum_{n=C}^{m-1}{[\ip{\L}_C]}_{n}Q_{\up}(n)+\hbar^2\sum_{n=C}^{m-1}Q_{\up}(n) -\Big[{[\ip{\L}_C]}_{C}+\hbar \sum_{n=C}^{m-1}Q_{\up}(n)\Big]^2,
\eea
The value of ${[\ip{\L^2}_C]}_{C}$ is just the square of the spinlabor cost for the situation where the memory is in the spin-up state, i.e. $(C\hbar)^2$, multiplied by the probability that it occurs, i.e. $p_\up$, and so ${[\ip{\L^2}_C]}_{C}=(C\hbar)^2 p_\up$.
Recalling that ${[\ip{\L}_C]}_{C}=C\hbar p_\up$, we find the variance for the full erasure process, obtained in the $m\to\infty$ limit, is
\bea
    \text{Var}(\L)_C &=& (C\hbar)^2 p_\up + 2\hbar\sum_{n=C}^{\infty}{[\ip{\L}_C]}_{n}Q_{\up}(n)+\hbar^2\sum_{n=C}^{\infty}Q_{\up}(n)-\Big[C\hbar p_\up+\hbar \sum_{n=C}^{\infty}Q_{\up}(n)\Big]^2  \notag\\
      &=& (C\hbar)^2 (p_\up-p_\up^2) + 2\hbar\sum_{n=C}^{\infty}\Big({[\ip{\L}_C]}_{n}-C\hbar p_\up\Big)Q_{\up}(n)+\hbar^2\sum_{n=C}^{\infty}Q_{\up}(n)-\hbar^2\Big[\sum_{n=C}^{\infty}Q_{\up}(n)\Big]^2,
\eea
and making use of \eqr{eqn:ap average spinlabor m = sum Q} this becomes
\bea
    \text{Var}(\L)_C &=& (C\hbar)^2 (p_\up-p_\up^2) + 2\hbar^2\sum_{n=C}^{\infty}\sum_{k=C}^{n-1}Q_{\up}(k)Q_{\up}(n)+\hbar^2\sum_{n=C}^{\infty}Q_{\up}(n)-\hbar^2\Big[\sum_{n=C}^{\infty}Q_{\up}(n)\Big]^2.
\eea
The first term on the right is the variance in the spinlabor cost for the CNOT steps before the first equilibration step, and the remaining terms constitute the variance in the cost for the CNOT steps that follow it; the fact that these contributions add to give the total variance is consistent with the fact that these two parts of the erasure process are statistically independent.
\end{widetext}

\bibliography{paper_2}

\end{document}